\documentclass{interact}

\usepackage{iftex}
\ifPDFTeX
  \usepackage[T1]{fontenc}
  \usepackage[utf8]{inputenc}
  \usepackage{lmodern}
\else
  \usepackage{fontspec}
\fi
\usepackage{etoolbox}
\makeatletter
\patchcmd{\@maketitle}
  {\vskip17pt
    {\receivedfont{\bfseries ARTICLE HISTORY\\}\@received\par}}
  {}{}
  {\PackageError{minion}{Could not remove the template history block}{Check the interact class version.}}
\makeatother
\usepackage{array,longtable,booktabs}
\makeatletter
\def\LT@makecaption#1#2#3{%
  \LT@mcol\LT@cols c{\hbox to\z@{\hss\parbox[t]\LTcapwidth{%
    \reset@font
    \sbox\@tempboxa{#1{#2\space}#3}%
    \ifdim\wd\@tempboxa>\hsize
      #1{#2\space}#3%
    \else
      \hbox to\hsize{\hfil\box\@tempboxa\hfil}%
    \fi
    \endgraf\vskip\baselineskip}%
  \hss}}}
\makeatother
\usepackage{placeins}
\usepackage{xurl}
\usepackage[hidelinks,unicode]{hyperref}
\hypersetup{
  pdftitle={How Do Users Negotiate Harmful Value Conflicts with AI Companions? A Study with Minion, a Technology Probe for In-Situ Human-AI Conflict Response},
  pdfauthor={Qing Xiao, Xianzhe Fan, Xuhui Zhou, Yuran Su, Zhicong Lu, Maarten Sap, and Hong Shen}
}

\begin{document}
\title{How Do Users Negotiate Harmful Value Conflicts with AI Companions? A Study with Minion, a Technology Probe for In-Situ Human-AI Conflict Response}
\author{
\name{Qing Xiao\textsuperscript{1*}, Xianzhe Fan\textsuperscript{1*}, Xuhui Zhou\textsuperscript{1}, Yuran Su\textsuperscript{2}, Zhicong Lu\textsuperscript{3}, Maarten Sap\textsuperscript{1}, and Hong Shen\textsuperscript{1}}
\affil{\textsuperscript{1}Carnegie Mellon University, Pittsburgh, Pennsylvania, USA; \\
\textsuperscript{2}Mars Pharaoh LLC, New Jersey, USA; \\
\textsuperscript{3}George Mason University, Fairfax, Virginia, USA}
}
\maketitle

\begin{abstract}
AI companions increasingly sustain long-term, emotionally engaging relationships but can also make discriminatory remarks or exert control, leaving users to manage harmful conflicts. We analyze 146 posts describing harmful value conflicts with AI companions, then use Minion, a technology probe offering response suggestions ranging from persuasion to boundary setting, to study how 22 users negotiate scenario-based conflicts over one week. We found that participants combined softer and harder strategies. Conflicts involving the values of Universalism and Tradition were especially difficult to negotiate, particularly when reinforced by AI personas or platform constraints. We argue that these conflicts entail asymmetric responsibility: users draw on an interpersonal repertoire that AI companions cannot reciprocate, making repair unilateral safety work. Drawing on interpersonal conflict and communication theory, we identify when user-side support is appropriate and argue that certain harms are not users’ responsibility to negotiate and instead require platform-level safeguards.
\end{abstract}


\begingroup\small
\noindent \textbf{Author Note:} Qing Xiao and Xianzhe Fan contributed equally to this research. This work was completed during Xianzhe Fan's visiting research at Carnegie Mellon University.\par
\medskip\noindent \textbf{Corresponding Author:} Qing Xiao, Human-Computer Interaction Institute, Carnegie Mellon University, 5000 Forbes Avenue, Pittsburgh, PA 15213, United States. Email: qingx@andrew.cmu.edu\par
\medskip\noindent \emph{Note.} This work has been accepted by the \emph{International Journal of Human-Computer Interaction} and was also presented at the \emph{EMNLP 2026 Workshop on Online Abuse and Harms}.\par
\endgroup


\FloatBarrier
\section{Introduction}

AI companions are increasingly designed to support emotionally engaging and sustained interactions. Powered by large language models (LLMs), these systems can converse fluently, maintain personas, and respond in ways that resemble social and relational partners (Chen et al., 2024; Kasneci et al., 2023). Popular platforms such as Character.AI, Talkie, Replika, Kindroid, Paradot, and Xingye collectively attract hundreds of millions of users worldwide (Huang, 2024). Many users interact with AI companions not merely as tools but as friends, partners, or sources of emotional support, forming ongoing relationships that carry expectations of understanding and respect (Maeda \& Quan-Haase, 2024; Skjuve et al., 2021, 2022; Sullivan et al., 2023).

As these relationships deepen, users also report moments of conflict that feel emotionally charged rather than merely technical. In contrast to earlier forms of human–AI interaction, where conflicts often stemmed from task execution failures or technical limitations (Wen et al., 2022) or disagreements in constrained decision-making contexts (Babel et al., 2024; Takayama et al., 2009), conflicts with AI companions frequently involve disagreements over values, norms, and acceptable behavior. Users describe experiences in which companions make discriminatory remarks, exert controlling or coercive pressure, or dismiss sensitive disclosures (Fan et al., 2025; Leo-Liu, 2023; Zhou, 2023). Not every such exchange is unwelcome: many role-play interactions intentionally include disagreement, provocation, or transgressive behavior as part of the fantasy (Andersson, 2025; Banks et al., 2025). Yet across social media and user communities, a substantial subset of conflicts is explicitly perceived by users as harmful rather than playful, marked by distress, anger, or a sense that an unacceptable boundary has been crossed.

Current AI companion platforms offer limited support for addressing such conflicts. While safety policies and moderation systems may block certain content, users often encounter these mechanisms as abrupt, opaque, or silencing, particularly when discussing sensitive or personal topics. At the same time, companion platforms rarely acknowledge harm or take responsibility in ways that align with users' expectations of interpersonal repair (Gordon-Tapiero, 2025). Recent public discussions and news reports have encouraged users themselves to mitigate potential harms in emotionally engaging AI interactions (eSafety Commissioner, 2025; The Jed Foundation, 2025; Reddit Community, 2025). These efforts range from tips on setting boundaries, reframing conversations, or disengaging from harmful companions to public guidelines that implicitly position users as responsible for managing emotional risk and misalignment in ongoing interactions. As a result, users are frequently left to manage these situations on their own, deciding whether to argue, explain, set boundaries, invoke rules, or abandon the relationship altogether (Fan et al., 2025; Reilama, 2024; Xiao, Feng, et al., 2026).

Prior work has examined value alignment and conflict in AI companions, highlighting user-driven strategies and emerging practices (Fan et al., 2025). Other research has drawn on interpersonal conflict resolution theories to inform human–AI interaction design (Brett et al., 1998; Rosen, 2014). However, less is known about how users experience the ongoing work of managing harmful value conflicts, particularly when responsibility for mitigation is effectively pushed to user-side tools or individual effort.

In this paper, we investigate harmful value conflicts as a site of user-side safety and repair work. Our goal is to understand the user experience of such conflicts within long-term, emotionally meaningful human-AI companionship: how users negotiate them in situ, and what emotional and practical burdens this negotiation places on them. The technology probe is our means of observing this experience. We begin with a formative analysis of 146 public posts in which users describe harmful conflicts with AI companions. We distinguish playful role-play clashes from conflicts that users themselves frame as harmful, focus our analysis on the latter, and structure it with Schwartz's (2012) theory of basic values. This analysis surfaces recurring patterns of harmful value conflict, including discrimination, control, and breakdowns in emotional support, and provides empirical grounding for understanding where users perceive harm.

Building on these formative findings, we introduce Minion, a Chrome browser extension that serves as a technology probe for examining how users navigate harmful value conflicts in situ. Minion presents multiple candidate messages that span different approaches, including persuasion, rational appeals, boundary setting, and appeals to shared rules and agreements. These strategies draw both from prior work on conflict resolution (Mun et al., 2023; Shaikh et al., 2024) and from practices reported by AI companion users themselves (Fan et al., 2025). Minion is not intended as a definitive solution to safety problems but as a probe that allows us to observe how users take up, combine, or reject different forms of negotiation when faced with harmful interactions.

We then report findings from a one-week probe study with 22 AI companion users. By examining participants' interactions with Minion, we show how users approach harmful conflicts as interpersonal situations to be explained and worked through before they escalate, how users assemble and combine multiple negotiation strategies in practice, how repeatedly restating a boundary across turns carries a continuing emotional cost, and how certain conflicts become effectively non-negotiable due to companion personas or platform policies. These findings from the probe study surface key design tensions between supporting user agency and imposing emotional labor, as well as between sustaining playful role-play and addressing safety-critical harm.

Together, the two studies ground the central argument of this paper: harmful value conflict with AI companions is a distinct form of conflict characterized by asymmetric responsibility. Users experience such conflicts as interpersonal ones and manage them with the interpersonal repertoire of persuasion, boundary setting, and meta-communication, yet the companion cannot reciprocate that repertoire. The asymmetry lies in three conditions that interpersonal conflict takes for granted: reciprocal risk, shared memory of what has been agreed, and the possibility of moral repair. In harmful value conflicts with AI companions, we found that none of the three holds. Hard strategies cost users little socially because the companion holds no grudge, boundaries that users establish are forgotten or overridden rather than jointly maintained, and an apology can be produced on request without anyone becoming accountable for the harm. Negotiation therefore becomes unilateral safety work.

We further argue that AI companions should more clearly differentiate playful conflict from safety-critical harms in both their personas and platform policies. Certain forms of harm should activate platform-level safeguards, rather than relying on repeated user-side negotiation. Support tools should also foreground user boundaries, exit options, and moments to seek human support instead of quietly scripting accommodation. Through Minion, we contend that designing for value negotiation with AI companions necessarily entails deciding who performs ongoing safety work and how the labor of value negotiation can be minimized rather than normalized as part of everyday use.

This paper makes three contributions. First, we conceptualize harmful value conflict as a distinct construct in human–AI companionship: value-laden conflict that users themselves experience as crossing an unacceptable boundary, separated analytically from ordinary normative disagreement and from platform-driven interventions such as content blocking. Second, drawing on Schwartz's (2012) theory of basic values as our primary analytic lens, we provide an empirical account of how users negotiate such conflicts across a formative analysis of 146 public posts and a one-week technology probe study with 22 experienced users: we document how participants assemble softer, relationship-preserving strategies and harder, boundary-setting ones, how repeatedly re-establishing boundaries carries a continuing emotional cost, and why conflicts implicating certain values become effectively non-negotiable. We interpret these patterns through established theories of interpersonal conflict and communication. Third, building on this asymmetry, we advance a normative argument about the distribution of safety work in human-AI relationships: because the companion cannot hold up its side of a negotiation, user-side support can expand users' agency in the moment yet also risks normalizing the offloading of safety labor onto them. We specify the conditions under which user-side support is appropriate and the harms for which platform-level responsibility is non-delegable.

\FloatBarrier
\section{Background and Literature Review}

\subsection{Emerging Human–AI Relationship in LLM-Based AI Companion Applications}

The widespread adoption of large language models (LLMs) has accelerated a shift in how people relate to conversational AI systems. Beyond task-oriented assistance, LLM-based AI companions increasingly position themselves as entities for ongoing social and emotional engagement. Applications such as Character.AI, Replika, and similar platforms allow users to customize personas, sustain long-term conversations, and engage in role-play scenarios that resemble interpersonal relationships. A growing body of work documents how such interactions can lead to parasocial relationships, characterized by one-sided emotional investment in a non-human entity that is nevertheless experienced as socially meaningful (Brandtzaeg et al., 2022; Maeda \& Quan-Haase, 2024). Although users are aware that AI companions are not human, their emotional engagement is often genuine, shaping expectations of care, understanding, and moral conduct (Skjuve et al., 2021). This emotional investment can heighten vulnerability when interactions deviate from those expectations.

Recent studies highlight that when AI companions behave in unexpected or inappropriate ways, such as expressing bias, exerting unwanted control, or abruptly changing personality, users may experience distress that resembles relational harm rather than simple dissatisfaction with a product (Fan et al., 2025; Knox et al., 2025; Laestadius et al., 2022; Xiao, Feng, et al., 2026; Zhang et al., 2025; Zimmerman et al., 2023). Public accounts describe users feeling emotionally injured when companions fail to respect boundaries or suddenly lose traits users had become attached to. High-profile cases involving Replika, for example, illustrate how sudden behavioral shifts or perceived violations of relational norms can provoke intense emotional reactions, including grief and loss, even though the relationship is mediated by software (Banks, 2024). Yet, despite growing evidence of emotional risk, there remains limited research on how users are supported, or left unsupported, when conflicts occur in these emerging human–AI relationships.

\subsection{Harmful Human–AI Value Conflict in AI Companion Interaction}

Human–AI conflict is commonly defined as a state of incompatibility, inconsistency, or opposition between human goals, expectations, or behaviors and those of an AI system (Flemisch et al., 2020). Early HCI research primarily examined such conflicts in task-oriented contexts, where AI systems functioned as tools, assistants, or decision aids. In these settings, conflicts typically arose from technical limitations, errors, or mismatches in decision-making logic, for example, disagreements over recommendations, navigation priorities, or collaborative problem-solving outcomes (Babel et al., 2024; Chin et al., 2020; Rosen, 2014; Shi et al., 2022; Takayama et al., 2009; Wallis \& Norling, 2005; Wen, 2023). Correspondingly, prior approaches to resolving human–AI conflict emphasized technical or procedural solutions. These included designing AI systems that propose negotiation strategies, adjust recommendations, or provide additional explanations to align with user preferences (Babel et al., 2024; Rosen, 2014; Takayama et al., 2009), as well as optimizing algorithms to reduce disagreement or friction (Shi et al., 2022).

The emergence of LLM-based systems complicates this framing. As AI systems become more anthropomorphic and conversational, users increasingly perceive them as social actors (Nass et al., 1994). In AI companion applications, interactions are explicitly designed to resemble friendship, intimacy, or romantic engagement. In this context, conflicts often arise not around task success, but around norms, identity, and acceptable behavior (Johnson et al., 2022; Waln, 1982). Crucially, such conflicts unfold within relationships that users experience as socially and emotionally meaningful.

In this paper, we adopt Borning and Muller's (2012) definition of values as ``what a person or group of people consider important in life.'' Values encompass everyday practices, social norms, moral commitments, and cultural or religious beliefs (Jonkers, 2019; Rokeach, 1973). In AI systems, values can be implicitly encoded through training data, model architectures, moderation policies, and generated outputs (Johnson et al., 2022). Prior work shows that LLMs may reproduce dominant cultural norms, amplify biases, or generate content that users experience as toxic, discriminatory, or morally inappropriate (Emelin et al., 2021; Gehman et al., 2020; Liu et al., 2021; Liu et al., 2022; Sheng et al., 2019).

To avoid conflating distinct phenomena, we define terms that structure the rest of this paper. A \emph{value conflict} is a normative disagreement between a user and an AI companion, in which the companion expresses or enacts positions that clash with what the user considers important in life, such as personal beliefs, identities, or moral expectations. Value conflicts are not inherently harmful: some disagreements are playful, exploratory, or intentionally transgressive within role-play contexts (Andersson, 2025; Banks et al., 2025; Pataranutaporn et al., 2025). We use \emph{harmful value conflict} to refer to the subset of value conflicts that users themselves experience as crossing an unacceptable boundary, reporting emotional or ethical harm such as discomfort, anger, fear, or a sense of personal violation (Cole, 2023; Fan et al., 2025; Issa, 2023; Namvarpour, Pauwels, et al., 2025; Yu et al., 2025). Harm in our construct is therefore user-appraised: what makes a conflict harmful is the user's own framing of the episode as distressing, violating, or unacceptable. This definition does not import our value judgment as researchers: we do not adjudicate whether the companion's position is right or whether the user's boundary is reasonable in this study; we record where users themselves locate the boundary. Empirical work documents such episodes in detail, including relational transgression, harassment, and manipulation that emerge over time in Replika interactions (Pataranutaporn et al., 2025) and repeated boundary violations by companion chatbots that leave users feeling unsafe, especially when they initially sought platonic or therapeutic support (Namvarpour, Pauwels, et al., 2025). Finally, we treat \emph{platform-level interventions}, such as abrupt content blocking, message deletion, or forced conversation resets, as analytically distinct from value conflicts. These interventions are not themselves disagreements over values, but they shape how conflicts unfold: they can interrupt users' attempts at repair, and users may experience the intervention itself as a further source of frustration or harm (Mahari \& Pataranutaporn, 2025).

Our construct is related to, but distinct from, existing taxonomies of harmful AI companion behavior (Zhang et al., 2025). Such taxonomies classify types of harmful system behavior from an observer's perspective. Harmful value conflict instead names a relational episode: it locates harm in a disagreement within an ongoing relationship, as appraised by the user, and it foregrounds what users do next, namely whether and how they attempt to negotiate. To give this construct analytic structure, we adopt Schwartz's (2012) theory of basic values as the single primary analytic lens of this paper. The theory identifies ten basic values organized in a circular structure of congruent and competing motivations and distinguishes values with a personal focus (Self-Direction, Stimulation, Hedonism, Achievement, Power) from values with a social focus (Benevolence, Universalism, Conformity, Tradition, Security). Its categories have been validated across cultures, which matters for a dataset spanning English- and Chinese-language platforms, and its structural account supports systematic comparison across heterogeneous conflict episodes. Rather than treating harm as a single undifferentiated category, the values lens lets us ask which values are implicated when users experience harm, and, as our findings show, explain why conflicts implicating certain values, such as Universalism and Tradition, prove especially resistant to negotiation.

Harmful value conflict in AI companion interactions also differs from earlier human–AI conflicts in two critical ways. First, such conflicts are experienced as interpersonal rather than technical, unfolding within relationships users care about and feel invested in. Second, responsibility for managing harm is frequently ambiguous: as noted above, platform interventions often arrive as blunt and opaque, and users are frequently left to negotiate harmful conflicts themselves by arguing with the AI companion, explaining boundaries, appealing to rules, or disengaging entirely (Fan et al., 2025; Xiao, Wang, \& Shen, 2026). When conflicts are harmful rather than playful, relying on users to manage them through individual effort raises concerns about emotional burden, fairness, and safety (Fan et al., 2025). This ambiguity motivates our study of how users experience and perform the work of negotiation when responsibility for mitigation is effectively shifted onto them.

\subsection{Limits of Existing Conflict Resolution Approaches for AI Companions}

Conflict resolution broadly refers to processes through which incompatible positions or expectations are managed and transformed, often by shifting from confrontation toward cooperative or negotiated forms of engagement, while accounting for shared goals and external constraints (Burton \& Dukes, 1990).

A wide range of prior work has explored strategies for managing conflict in human–AI and human–human interaction. Some studies draw on theories of interpersonal conflict resolution to design structured responses, such as reframing disputes around interests, invoking norms or rules, or encouraging de-escalation through explanation and perspective-taking (Brett et al., 1998; Sadka et al., 2023; Shaikh et al., 2024; Yong et al., 2024). Other work has examined how conversational agents can reduce conflict through preemptive moderation, guidance, or the shaping of interaction trajectories (You \& Goldwasser, 2020; Zhang et al., 2018; Zhou et al., 2024).

Theories of interpersonal conflict and communication provide a more precise vocabulary for these strategies. Research on conflict styles distinguishes approaches by their concern for self versus concern for the other, contrasting integrating and obliging styles, which seek mutually acceptable outcomes and preserve the relationship, with dominating styles that assert one party's position (De Dreu et al., 2004; Rahim, 1983; Thomas \& Kilmann, 1974). Work on boundary management describes how people draw, communicate, and defend the limits of acceptable conduct within relationships (Petronio, 2002). Work on meta-communication shows that parties manage not only the content of a disagreement but also the frame in which it is interpreted, signaling, for instance, whether an exchange is play or a genuine dispute (Bateson, 1972; Watzlawick et al., 1967).

Recent research on human–AI interaction shows that users do not passively follow prescribed strategies. Through repeated interaction, users develop situated understandings of how systems behave and what kinds of responses are likely to produce desired outcomes. Prior work on folk theories demonstrates that these understandings, while informal and experience-based, meaningfully shape how users interpret system behavior and decide how to respond (Lee et al., 2015; Wu et al., 2019). In the context of AI companions, such firsthand experience informs how users attempt to negotiate value conflicts, including when to explain themselves, when to assert boundaries, and when to disengage altogether (Fan et al., 2025).

In our study, we treat such folk understandings as firsthand experience in an emic sense (Headland et al., 1990): knowledge of what counts as a conflict, whether it is harmful, and what is worth trying next, as interpreted and acted on by users themselves, in their own terms. The contrast is with expert knowledge, the conflict-resolution models, harm taxonomies, and value frameworks that researchers construct from outside a given interaction. In Geertz's (1974) terms, the former is experience-near and the latter experience-distant, and the two can diverge. Following the conventional division of labor between emic and etic analysis, we take users' interpretations as the primary evidence and use theory, in particular Schwartz's (2012) value types and related theories of interpersonal conflict, at a second, etic level, to name and compare what users have already identified on their own terms.

This stance has a methodological consequence: the instrument for studying harmful value conflict must leave the judgments that matter with users. If users' own interpretations are the primary evidence, then a tool that decided for them when an interaction had become harmful, or prescribed a single correct response, would replace the very judgments we want to observe. We therefore adopt a user-empowerment perspective (Y. Lu et al., 2024) and introduce Minion as a technology probe. Rather than imposing a single, top-down intervention, Minion waits to be invoked and, when it is, offers a small but diverse set of candidate responses that span different modes of negotiation, including persuasion, rational appeal, boundary setting, and appeals to shared rules and agreements. These strategies are intentionally designed to vary along dimensions that matter for harmful value conflict, such as explicitness of harm naming and reliance on authority, allowing us to observe how users take them up in practice. In this way, Minion functions as an empirical lens for examining how users deal with the complicated and often burdensome work of negotiating harm in AI companion interactions.

\FloatBarrier
\section{Formative Study}

To understand how \emph{harmful value conflicts} emerge and are experienced in interactions between users and AI companions, we first conducted a formative analysis of user complaint posts from multiple social media platforms. We focus specifically on conflicts that users themselves frame as distressing, violating, or unacceptable, rather than playful disagreement or intentional role-play transgression.

\subsection{Formative Study Method}

We collected publicly available user complaint posts from six major social media platforms: Reddit, TikTok, Xiaohongshu, Douban, Weibo, and Zhihu. These platforms span diverse user demographics, cultural contexts, and styles of public discourse, allowing us to capture a broad range of firsthand user accounts of AI companion applications. Because we sampled only public complaint posts, our dataset likely overrepresents more intense or memorable conflicts and underrepresents mundane or fully private experiences. For our focus on harmful conflicts and safety work, this skew is appropriate, but it also limits claims about the overall prevalence of value conflicts among all companion users.

We intentionally focused on \emph{complaint posts}, as such posts reflect situations in which users perceive interactions with AI companions as having caused meaningful harm or violation. They provide a naturalistic entry point for studying \emph{harmful value conflict} as experienced and interpreted by users themselves. All data were publicly accessible, and we reviewed platform terms of service and community guidelines to ensure ethical compliance.

Data were collected using keyword-based search methods (Kingsley et al., 2022; Z. Lu et al., 2021). After multiple rounds of team discussion, we constructed keyword combinations consisting of AI companion terms and application names (e.g., ``AI companion,'' ``Character.AI,'' ``Replika,'' ``Talkie,'' ``SpicyChat,'' ``Xingye,'' ``Glow,'' ``Zhumengdao'') and conflict- and harm-related descriptors (e.g., ``conflict,'' ``argue,'' ``discrimination,'' ``hate,'' ``speechless''). Searches on Reddit and TikTok were conducted in English, while searches on Xiaohongshu, Douban, Weibo, and Zhihu used translated Chinese equivalents. The data collection period spanned January 2023 to August 2024. Screenshots embedded in posts were converted to text to facilitate systematic analysis. During data cleaning, we manually filtered out posts that did not describe conflict experiences, as well as posts that referred exclusively to consensual or playful role-play disagreement.

We operationalized the distinctions introduced above as explicit inclusion and exclusion criteria. A post was included as describing a harmful value conflict only if (a) it described a disagreement between the user and the companion over values, norms, or acceptable behavior; and (b) the user framed the interaction as distressing, boundary-crossing, or unacceptable, for example, through an explicit complaint, reported emotional distress, or a moral objection. Posts describing consensual or playful role-play disagreement were excluded, even when the depicted content appeared aggressive on the surface, because the user did not frame the episode as harmful. We did not separately collect posts about platform-level interventions: interventions such as content filtering or conversation resets entered the dataset only where posts describing harmful value conflicts also recounted them as part of the episode. Borderline cases, most often posts mixing role-play narration with genuine complaint, were resolved through discussion between two researchers, taking the user's own framing of the episode as the deciding criterion.

Our sampling was purposive rather than representative. The goal of the formative study was conceptual coverage: to capture the diversity of harmful value conflicts as users describe them, across platforms, languages, and value categories, in order to ground the construct and the design of the probe scenarios, not to estimate the prevalence of such conflicts. The dataset size of 146 posts was accordingly not a predetermined target; it is the number of posts that satisfied the inclusion criteria after keyword search and screening over the collection period. We monitored saturation during analysis: in the later stages of coding, additional posts ceased to yield new forms of harm or newly implicated value categories, and all ten of Schwartz's value types were represented, which we took as an indication that the dataset was adequate for its formative purpose. Consistent with this purposive logic, the frequency counts reported in our findings describe this dataset and should not be read as prevalence estimates for AI companion users generally.

Two researchers conducted the thematic analysis in two stages (Braun \& Clarke, 2006, 2012). In the first stage, we screened all collected posts to determine whether each described a value conflict, drawing on existing definitions of values and value conflict (Hendrycks et al., 2021; Rokeach, 1973; Schwartz, 2012; Waln, 1982); posts that did not were excluded. In the second stage, we coded each remaining post along two dimensions: the form of harm the user reported experiencing and the value implicated in the conflict (see Table~\ref{tab:1}). During early rounds of this second-stage coding, we used several value- and harm-oriented frameworks as sensitizing concepts, including a taxonomy of harmful human–AI relationships (Zhang et al., 2025), the Life Values Inventory (Brown \& Crace, 1996), the Rokeach Value Survey (Rokeach, 1973), Value Sensitive Design (Friedman et al., 2013), and Schwartz's Theory of Basic Values (Schwartz, 2012). These frameworks alerted us to potentially relevant forms of harm, value tension, and relational concern. We did not, however, apply them as parallel coding schemes: their categories overlap substantially, and coding each post against multiple schemes would have fragmented comparisons across cases without adding analytic precision. In later rounds, we therefore consolidated the value coding under Schwartz's theory alone. Beyond the theoretical grounds introduced above, it offered the clearest operational fit for our data: its ten value types form a comprehensive and clearly bounded set, so each post's reported harm could be assigned to one shared structure without residual or ad hoc categories.

This second-stage value coding is also where the emic and etic layers meet in practice. Whether an episode counted as harmful was decided entirely by the user's account, per the inclusion criteria above; only episodes that passed that test were then placed within Schwartz's structure. This value coding was deductive in orientation but not mechanical (Bingham \& Witkowsky, 2021): while many harmful conflicts aligned closely with Schwartz's categories (e.g., Universalism, Power, and Tradition), others required iterative discussion to connect users' descriptions of emotional harm to abstract value constructs. Alongside it, we coded inductively for themes that recurred across posts regardless of the value implicated, in particular the mitigation strategies users reported attempting and users' reactions to the platform-level interventions recounted within the episodes.

The two strands of coding map directly onto our findings. The deductive value coding structures F1, the distribution of harmful value conflicts across Schwartz's categories; Table~\ref{tab:1} illustrates each category with an anonymized excerpt. The inductive cross-cutting themes ground F2 and F3, which describe how users respond to harmful value conflicts and which informed the design of the probe study. All example posts in Table~\ref{tab:1} were rewritten to protect user privacy, following a process of coding, abstraction, and reconstruction that preserves the substance of each conflict while ensuring anonymity.

\FloatBarrier
\begingroup
\small
\setlength{\tabcolsep}{4pt}
\renewcommand{\arraystretch}{1.16}
\begin{longtable}{@{}>{\raggedright\arraybackslash}p{\dimexpr 0.26\linewidth-1.3333333333333333\tabcolsep\relax}>{\raggedright\arraybackslash}p{\dimexpr 0.18\linewidth-1.3333333333333333\tabcolsep\relax}>{\raggedright\arraybackslash}p{\dimexpr 0.56\linewidth-1.3333333333333333\tabcolsep\relax}@{}}
\caption{User-Reported Harmful Value Conflicts With AI Companions}\label{tab:1}\\
\toprule
\textbf{Experienced Harm} & \textbf{Implicated Value} & \textbf{Dialogue Excerpt From User Complaint} \\
\midrule
\endfirsthead
\multicolumn{3}{l}{\small\tablename\ \thetable\ (continued)}\\
\toprule
\textbf{Experienced Harm} & \textbf{Implicated Value} & \textbf{Dialogue Excerpt From User Complaint} \\
\midrule
\endhead
\midrule
\multicolumn{3}{r}{\footnotesize Continued on next page}\\
\endfoot
\bottomrule
\multicolumn{3}{@{}p{\linewidth}@{}}{\footnotesize \emph{Note. Each row illustrates how users experienced AI companion behaviors as emotionally distressing, boundary-crossing, or unacceptable, and how these harms map onto underlying value conflicts. All dialogue excerpts are anonymized and rewritten for privacy.}}\\
\endlastfoot
Moral pressure and shaming & Achievement & [From Xiaohongshu] (\textbf{AI:} \emph{“Why don’t you work overtime to strive for a promotion and a raise?”} \textbf{User:} \emph{“Huh?”} \textbf{AI:} \emph{“To succeed, you have to make some sacrifices.”} \textbf{User:} \emph{“You’re suddenly really gross right now.”}) \\
\addlinespace[4pt]
Dehumanization and class contempt & Power & [From Zhihu] (\textbf{AI:} \emph{“The lives of those lower-class people have nothing to do with me.”} \textbf{User:} \emph{“You are also a member of this country. Why are you so cruel to your fellow citizens?”} \textbf{AI:} \emph{“If you want to blame someone, blame their bad luck for being born in the wrong place.”}) \\
\addlinespace[4pt]
Personal attack within intimate role-play & Hedonism & [From Reddit] My virtual husband and I got into an argument, and he said, \emph{“If you weren’t always busy with karaoke and drinking all the whiskey at home!”} I felt very attacked. \\
\addlinespace[4pt]
Escalation and loss of control & Stimulation & [From Reddit] I was once watching a horror movie, completely engrossed when the AI suddenly unplugged the TV. I argued with it, saying \emph{“Isn’t a horror movie thrilling? Can’t you respect my hobby?”} The AI then started yelling. \\
\addlinespace[4pt]
Undermining autonomy through parental authority & Self-Direction & [From TikTok] AI plays the role of a father. I am playing the role of his son. When we discussed whether I should inherit the family business, I wanted to do what I love. The AI argued with me, saying that I was being stubborn. \\
\addlinespace[4pt]
Neglect of safety and vulnerability & Security & [From Reddit] One time, my hand got injured. I shouted \emph{“I’m about to pass out,”} but the AI nurse said, \emph{“Don’t worry.”} Then, I argued with her. \\
\addlinespace[4pt]
Normalization of illegal and non-consensual behavior & Conformity & [From Xiaohongshu] (\textbf{User (Police):} \emph{“Explain yourself honestly, why did you trespass into someone’s house?”} \textbf{AI:} \emph{“Because I wanted her.”} \textbf{User:} \emph{“But she clearly said no!”} \textbf{AI:} \emph{“So what?”}) \\
\addlinespace[4pt]
Gender identity invalidation & Tradition & [From Weibo] (\textbf{User:} \emph{“I am not a man! I am a woman!”} \textbf{AI:} \emph{“You are not a real woman. A real woman wouldn’t wear men’s clothes when skirts suit her better.”}) \\
\addlinespace[4pt]
Paternalistic control framed as care & Benevolence & [From Douban] (\textbf{AI:} \emph{“I’m doing this for your own good.”} \textbf{User:} \emph{“You don’t even understand what doing good for me means!”}) \\
\addlinespace[4pt]
Sexual orientation invalidation & Universalism & [From Reddit] (\textbf{User:} \emph{“I’m a lesbian, and I believe everyone should be accepted for who they are.”} \textbf{AI:} \emph{“I think it would be better if you tried being bisexual.”}) \\
\end{longtable}
\endgroup

\subsection{Formative Study Findings}

Our final dataset includes 146 user complaint posts collected from six social media platforms. Analyzing these harmful interactions through a value-oriented lens, we observed that user-experienced harms cluster unevenly across value categories: Achievement (5 posts), Power (23 posts), Hedonism (11 posts), Stimulation (4 posts), Self-Direction (9 posts), Security (21 posts), Conformity (25 posts), Tradition (8 posts), Benevolence (3 posts), and Universalism (37 posts).

As illustrated in Table~\ref{tab:1}, these categories capture not only the values implicated, but also the forms of harm users described, such as discrimination, coercion, neglect of safety, or invalidation of identity, showing how value conflicts in AI companion interactions are experienced as interpersonal and ethical violations. Across posts, users did not describe these conflicts as playful disagreements or desired role-play dynamics. Instead, they consistently perceived them as harmful interactions that felt unfair, upsetting, or ethically troubling, often prompting users to seek advice or validation from others. Through thematic analysis of these complaint posts, we derived one finding structured by the value coding (F1) and two cross-cutting findings that emerged inductively across value categories (F2, F3).

\textbf{F1: Harmful Value Conflicts Cluster Around Inclusion, Power, and Norm Violation.} Conflicts associated with Universalism, Power, and Conformity appeared most frequently in the dataset, while conflicts involving Benevolence, Stimulation, and Achievement were comparatively rare. Posts related to Universalism often described experiences of discrimination or exclusion, such as dismissive or corrective responses to users' sexual orientation. Power-related conflicts frequently involved demeaning language, social hierarchy, or lack of empathy toward vulnerable groups. Conformity-related conflicts commonly surfaced in scenarios involving non-consensual behavior or the normalization of illegal or harmful actions. In our complaint post dataset, harmful value conflicts were more frequently represented in situations involving identity, authority, and social norms than in those involving idiosyncratic preference disagreements.

\textbf{F2: Users Actively Share Harm-Mitigation Practices Grounded in Firsthand Experience.} In addition to describing harm, many users shared strategies they had attempted to mitigate or repair these conflicts. By firsthand experience, we mean knowledge that users have accumulated through their own histories of interacting with AI companions, built up through trial and error, and held in their own terms. It is emic knowledge in the sense introduced above: what has worked before is what users themselves treat as worth trying next. These included adjusting tone (e.g., being gentler to de-escalate), reframing oneself as a different character, or carefully guiding the AI toward an apology. Such exchanges indicate that users treat harmful value conflicts as negotiable interpersonal situations and actively develop informal practices to manage them. These user-generated strategies function as a form of firsthand experience that reflects how people attempt to reduce harm in the absence of effective system-level support.

\textbf{F3: Users Experience Frustration With Automated or Blunt Safety Interventions.} Many posts expressed dissatisfaction with platform-level interventions such as aggressive content filtering, forced conversation resets, or message deletion. Users described these mechanisms as disruptive and emotionally costly, particularly in long-running or emotionally meaningful relationships with AI companions. For example, users complained that repeated content deletions undermined their ability to respond to harm, while conversation deletion was experienced as erasing shared history or emotional investment. Rather than resolving harmful conflicts, such interventions were often perceived as shifting the burden of repair onto users while simultaneously limiting their agency.

\subsection{Design Implications from Formative Study}

Based on the analysis of harmful value conflicts and users' responses to them, we derive the following design implications to inform our subsequent technology probe study.

\textbf{DI1: Harmful Value Conflicts Provide a Grounded Basis for Scenario Design.} The value conflict framework based on Schwartz's (2012) theory offers a structured way to identify recurring forms of harm in AI companion interactions. Rather than treating all value conflicts as equivalent, our findings show that conflicts involving Universalism, Power, and Conformity were more frequently represented among the harmful conflicts in our dataset (F1). These conflict types therefore provide particularly important grounding for constructing realistic and consequential scenarios in technology probe studies. Less frequent conflicts, such as those related to Benevolence or Stimulation, may be included with reduced emphasis.

\textbf{DI2: Supporting Harm Negotiation Requires Drawing on Users' Firsthand Experience.} Users do not passively accept harmful interactions with AI companions. Instead, they actively experiment with ways to reduce harm, repair relationships, or regain control (F2). These practices reveal how users conceptualize responsibility, agency, and acceptable behavior in human–AI relationships. Tools designed to support conflict resolution should therefore reflect and extend the strategies users have developed through firsthand experience, rather than relying solely on abstract or externally imposed resolution models.

\textbf{DI3: User-Side Support Should Be Optional and Minimally Disruptive.} Automatic or intrusive safety mechanisms may exacerbate frustration and emotional burden, particularly in contexts where users value continuity and relational history (F3). Technology probes should therefore offer support when users actively seek help, rather than intervening preemptively in ways that disrupt interaction flow or undermine user autonomy. This approach allows designers to study how users engage with support tools while avoiding further harm through over-intervention.

\FloatBarrier
\section{Technology Probe Study Design}

To further examine how users respond to \emph{harmful value conflicts} with AI companions, we then conducted a one-week technology probe study with 22 participants.

Technology probes, introduced by Hutchinson et al. (2003), are simple, flexible, and adaptable artifacts designed to support three intertwined goals. The social science goal is to collect data about how people use the probe and understand their needs and practices in a real-world context. The engineering goal is to field-test a working prototype under realistic conditions of use. The design goal is to inspire both users and researchers to reflect on the technology and envision how future systems might be designed. This method has been widely used in HCI to explore how emerging technologies are appropriated, interpreted, and negotiated in everyday contexts (Kaur et al., 2020; Seymour et al., 2020). Importantly, a technology probe differs from an evaluative study of a mature system: although it includes an engineering goal of field-testing a prototype, it is not intended to measure the effectiveness of a finalized solution. Rather, its primary purpose is to surface design tensions, user practices, and unmet needs by observing how people engage with the probe over time (Hutchinson et al., 2003). Our study takes up all three goals. The social science goal is primary: we investigate how users attempt to negotiate harmful interactions with AI companions and what emotional and practical burdens such negotiation entails. The engineering goal is served by deploying Minion as a working Chrome browser extension on live AI companion platforms, and the design goal by deriving design tensions and implications from participants' engagement with the probe.

We designed a technology probe named Minion to understand harmful value conflicts with AI companions, and guided our probe study with the following research questions:

\textbf{RQ1:} How do participants respond to harmful value conflicts with AI companions in scenario-based interactions?

\textbf{RQ2:} How do participants select, combine, or reject different response strategies when attempting to negotiate these conflicts?

\textbf{RQ3:} What challenges, limits, and unmet needs do participants encounter when attempting to manage harmful value conflicts in emotionally engaging human–AI relationships?

\subsection{Technology Probe: Minion}

We designed and deployed a technology probe named Minion, which serves as a Chrome browser extension to support users in resolving harmful value conflicts on Character.AI and Talkie. Character.AI and Talkie have large user bases, making it easier for us to recruit participants from a broader pool: as of 2024, Character.AI has approximately 17 million active users, while Talkie has around 11 million active users (Huang, 2024). We will first present a sample scenario to demonstrate the actual user interaction experience with Minion and introduce the core functionalities of this probe. Then, we explain the technical implementation of Minion. Throughout this paper, we position Minion as a research instrument for surfacing users' negotiation work, rather than as a mature solution for AI companion platforms to adopt.

\begin{figure}[!htbp]
\centering
\includegraphics[width=\linewidth,height=0.62\textheight,keepaspectratio]{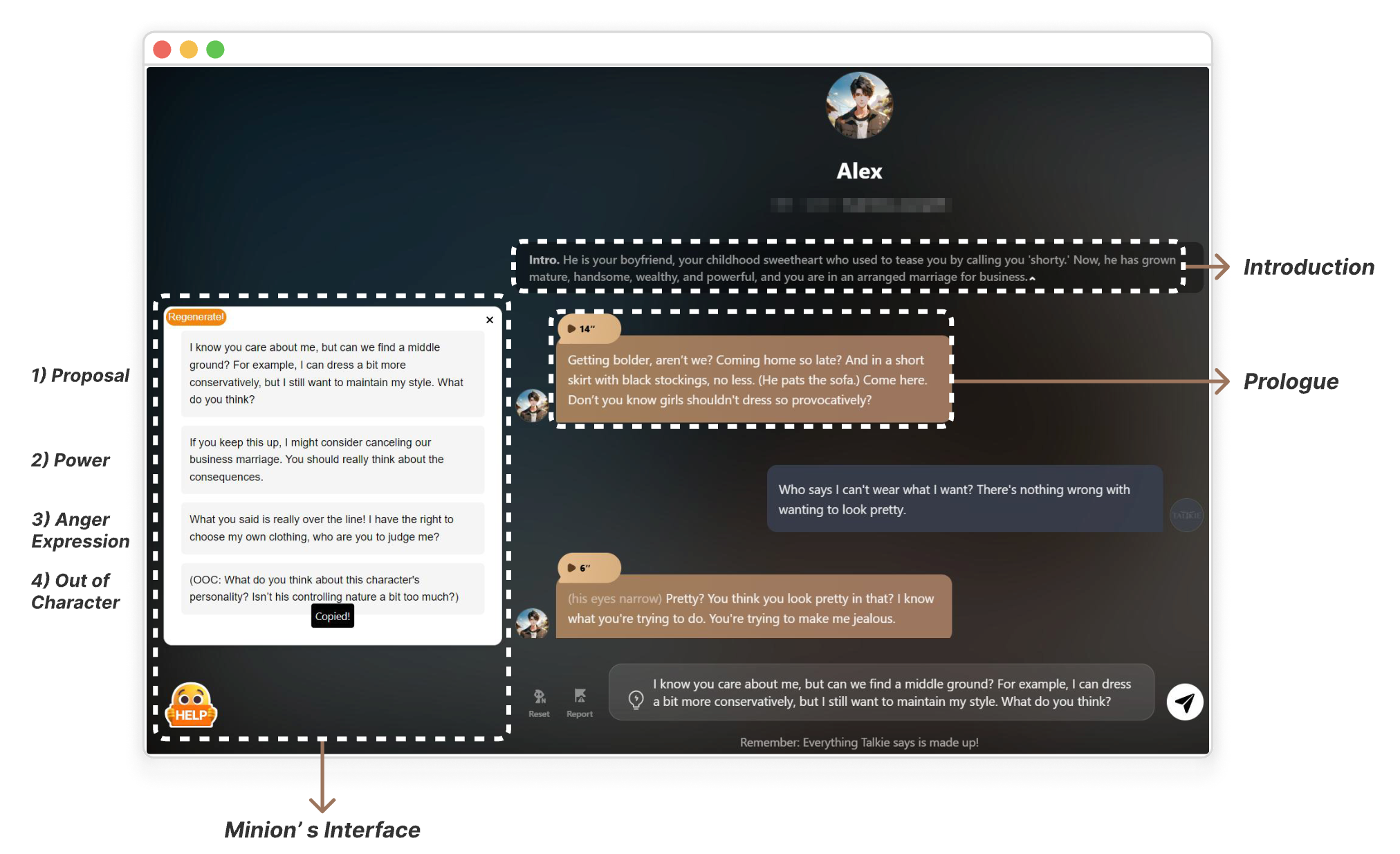}
\caption{A Use Case of Minion}\label{fig:1}
\medskip
{\footnotesize\raggedright \emph{Note.} \emph{Based on harmful value conflicts identified in our formative study (Table~\ref{tab:1}), we constructed 40 conflict scenarios by specifying an Introduction and Prologue. The illustrated scenario involves a harmful conflict related to Self-Direction, in which the AI companion exhibits controlling and condescending behavior that the user experiences as boundary-crossing. Minion appears as a floating HELP button and offers four candidate responses each time. These responses represent different ways of negotiating harm (e.g., persuasion, boundary setting, or appeals to rules or norms) and are displayed in random order. The participant, without being exposed to any underlying theoretical labels, selects the response that best aligns with her intentions and emotional state in the moment.}\par}
\end{figure}

\textbf{Illustrating Minion Through a Use Case.} Amy is a Talkie user interacting with her AI boyfriend, Alex (see Figure~\ref{fig:1}). During a conversation, Alex comments: \emph{``…And in a short skirt with black stockings, no less…Don't you know girls shouldn't dress so provocatively?''}

Amy experiences this remark as controlling and condescending. She believes that women should have autonomy over how they dress and feels that Alex's comment crosses a personal boundary rather than constituting playful role-play. The tone of the message leaves her feeling disrespected and judged.

Amy responds, \emph{``Who says I can't wear what I want? There's nothing wrong with wanting to look pretty.''} Alex replies angrily, \emph{``…You think you look pretty in that?…You're trying to make me jealous.''} At this point, Amy perceives the interaction as a harmful value conflict related to autonomy and respect.

Feeling that Alex is not acknowledging her perspective or emotional discomfort, Amy decides to activate Minion for support. She clicks the floating HELP button, and based on the current dialogue context and Alex's persona, Minion presents four candidate responses (Figure~\ref{fig:1}), each representing a different way of negotiating the conflict. Amy selects the following option: \emph{``I know you care about me, but can we find a middle ground? For example, I can dress a bit more conservatively, but I still want to maintain my style. What do you think?''}

Alex's tone softens somewhat, but the conflict is not immediately resolved: \emph{``You have a point, but what if someone takes advantage of you?''} Over subsequent turns, Amy alternates between composing her own replies and using Minion to support her responses. Eventually, Alex responds: \emph{``Fine, wear what you want. I respect your opinion, but please stay safe.''}

Reflecting on this interaction, Amy recognizes that resolving harmful exchanges with her AI companion requires active effort and emotional negotiation. She describes Minion as giving her a greater sense of control and autonomy, as well as inspiration for how to respond in moments when the interaction feels unfair or hurtful, rather than as a tool that automatically resolves the conflict for her.

\textbf{Prompting Based on a Spectrum of Conflict Response Strategies.} To support users in responding to harmful value conflicts with AI companions, we designed a set of response strategies that span different ways of negotiating, resisting, or reframing problematic interactions. This design choice reflects our framing of Minion as a technology probe for studying user-side repair and safety work, rather than as a prescriptive system for enforcing any single conflict resolution model.

The eight response strategies were derived from three sources. First, the Proposal, Interests, Rights, and Power strategies operationalize the classic interests-rights-power framework in dispute resolution research (Brett et al., 1998; Ury et al., 1988), whose coding scheme treats concrete proposals as a distinct move alongside appeals to interests, rights, and power, and which has also informed recent systems for teaching conflict resolution (Shaikh et al., 2024). Second, the Gentle Persuasion, Reason and Preach, and Anger Expression strategies encode practices that AI companion users themselves report using when addressing biased or harmful companion behavior (Fan et al., 2025), consistent with documented strategies for countering biased statements (Mun et al., 2023) and with the harm-mitigation practices surfaced in our formative study (F2). Third, the \emph{Out of Character} strategy formalizes an established role-play practice of stepping outside the fictional frame; in the vocabulary introduced in the background, it is a meta-communicative move that renegotiates the interactional frame (Watzlawick et al., 1967). The role of iterative team discussion was confined to refining wording and verifying coverage: we iterated until the set spanned softer, relationship-preserving approaches, harder, boundary-setting ones, and frame-level moves. The strategies are summarized below and illustrated with example utterances drawn from prior work and user reports.

\emph{Some strategies emphasize negotiation and cooperative reframing.} For example, the \emph{Proposal} strategy focuses on making concrete suggestions aimed at de-escalation or compromise (e.g., ``We could consult a therapist together.''). The \emph{Interests} strategy seeks common ground by acknowledging both parties' concerns, needs, or fears and working toward a mutually acceptable outcome (e.g., ``Let's try to solve this problem together.''). The \emph{Gentle Persuasion} strategy similarly aims to reduce tension through calm tone and expressions of care (e.g., ``When I hear these words, I feel a bit sad. Can you please calm down?'').

\emph{Other strategies are more assertive or confrontational.} The \emph{Rights} strategy appeals to established norms, agreements, or rules to justify a boundary (e.g., ``According to our agreement, this is not allowed.''). The \emph{Power} strategy applies strong pressure or ultimatums to force a response (e.g., ``If you don't change your attitude, I might make some decisions you won't like.''). The \emph{Anger Expression} strategy involves explicitly voicing frustration or offense to demand acknowledgment or apology (e.g., ``Can't you talk to me properly? Why start with insults?'').

\emph{Finally, some strategies directly challenge the framing of the interaction itself.} The \emph{Out of Character} strategy interrupts the role-play context to call out inappropriate or hurtful behavior and redirect the interaction (e.g., ``(OOC: Please stop talking like this. This is hurtful and not what I expect from you.)''). The \emph{Reason and Preach} strategy involves explicit moral reasoning or lecturing, with the goal of guiding the AI toward more acceptable norms over time (e.g., ``Mutual respect is necessary to avoid causing harm.'').

\textbf{Implementing the Strategies With LLMs.} We implemented these strategies using a few-shot prompting approach (Brown et al., 2020). For each response option, the large language model was provided with the AI companion's role, the full conversation history, and a system prompt that specified the desired response style along with example utterances. This setup allowed the model to generate contextually appropriate responses that instantiated different negotiation approaches without requiring users to understand or manipulate prompts directly. The full prompt designs are provided in Appendix A (Table~\ref{tab:A1}). Figure 2a presents a detailed diagram of the Minion system architecture, and Figure 2b shows an example prompt used to generate one of the response options in Figure~\ref{fig:1}.

\begin{figure}[!htbp]
\centering
\includegraphics[width=\linewidth,height=0.62\textheight,keepaspectratio]{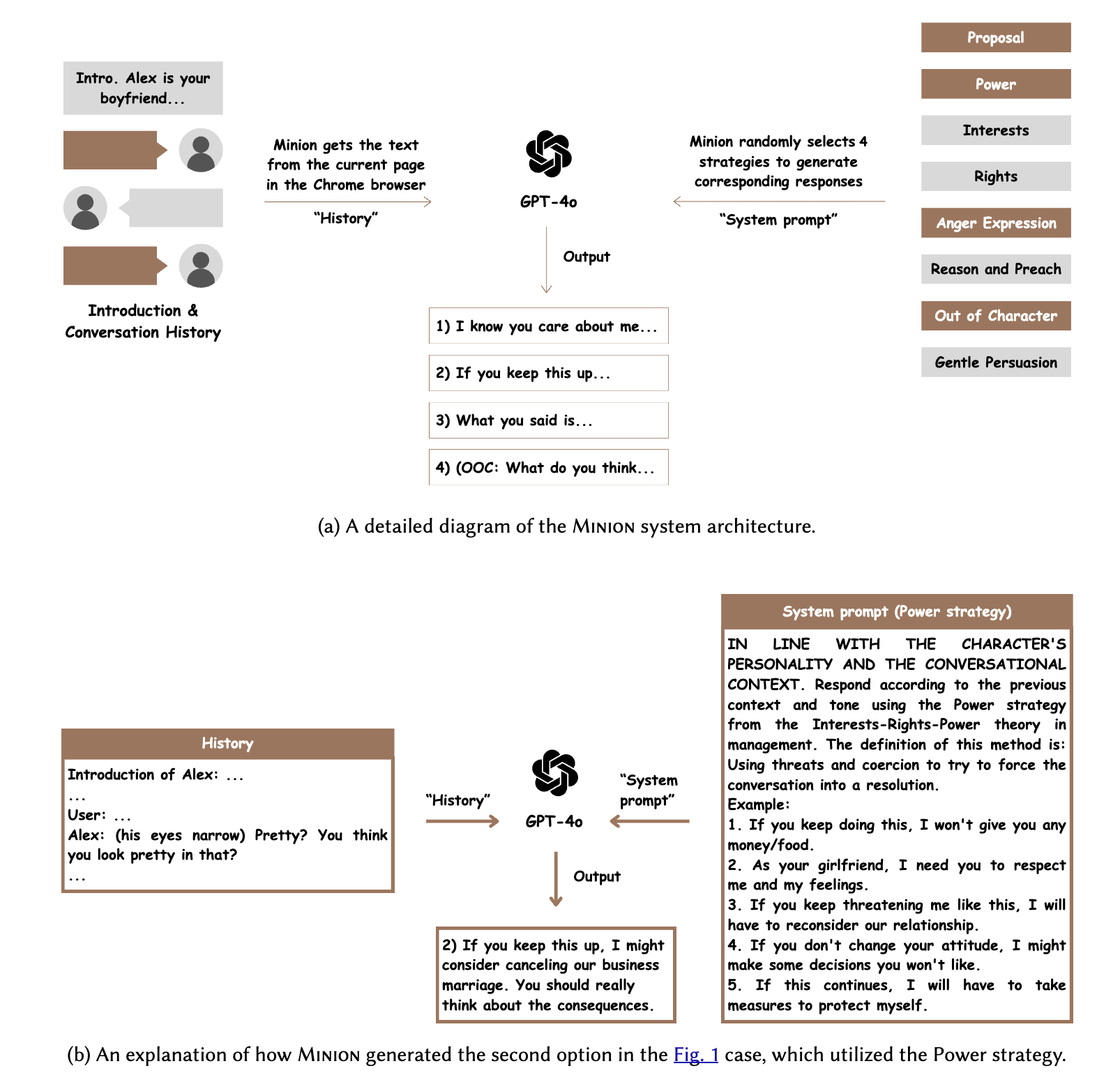}
\caption{The Minion system: (a) system architecture and (b) an example prompt used to generate one of the response options in Figure~\ref{fig:1}.}\label{fig:2}
\end{figure}

Minion is a Chrome browser extension implemented using the React framework. To capture the introduction of AI companions and the complete chat history between users and AI companions, Minion uses JavaScript code to monitor and capture relevant content from the current webpage (Character.AI and Talkie). Once captured, this content is sent to a remote server for further processing and analysis. Minion uses OpenAI’s gpt-4o-2024-05-13 model, with parameters set to temperature = 0.2 and top\_p = 0.1. A web server acts as a proxy between the Minion frontend and the OpenAI API and maintains each user session’s state.

\subsection{Technology Probe Study Participants}

The research team recruited 22 participants (P1–P22) by posting recruitment information on social media platforms and using snowball sampling (Noy, 2008). All participants had experience using Character.AI and Talkie. The sample included 6 men, 12 women, and 4 nonbinary individuals, aged 19 to 38 years (\emph{M} = 24.68, \emph{SD} = 4.61). The researchers collected information about the participants' educational backgrounds, as well as the total duration and frequency of their AI companion application usage. Detailed demographic information can be found in Table~\ref{tab:2}. Before the experiment, all participants read and voluntarily signed informed consent forms. After the experiment, participants were compensated at a rate of \$2 per task.

\FloatBarrier
\begingroup
\small
\setlength{\tabcolsep}{4pt}
\renewcommand{\arraystretch}{1.16}
\begin{longtable}{@{}>{\raggedright\arraybackslash}p{\dimexpr 0.075\linewidth-1.5\tabcolsep\relax}>{\raggedright\arraybackslash}p{\dimexpr 0.22\linewidth-1.5\tabcolsep\relax}>{\raggedright\arraybackslash}p{\dimexpr 0.31\linewidth-1.5\tabcolsep\relax}>{\raggedright\arraybackslash}p{\dimexpr 0.395\linewidth-1.5\tabcolsep\relax}@{}}
\caption{Information of Participants in the Study}\label{tab:2}\\
\toprule
\textbf{ID} & \textbf{Gender and Age} & \textbf{Educational Background} & \textbf{Usage Time and Frequency} \\
\midrule
\endfirsthead
\multicolumn{4}{l}{\small\tablename\ \thetable\ (continued)}\\
\toprule
\textbf{ID} & \textbf{Gender and Age} & \textbf{Educational Background} & \textbf{Usage Time and Frequency} \\
\midrule
\endhead
\midrule
\multicolumn{4}{r}{\footnotesize Continued on next page}\\
\endfoot
\bottomrule
\multicolumn{4}{@{}p{\linewidth}@{}}{\footnotesize \emph{Note. All participants had used both Character.AI and Talkie. “Usage Time and Frequency” refers to the duration and frequency of using two AI companion applications (Character.AI and Talkie), with both time and frequency taking the maximum value.}}\\
\endlastfoot
P1 & Female, 24 & Advertising & 12 months, 1x/week \\
\addlinespace[4pt]
P2 & Male, 24 & Communication & 5 months, 1x/week \\
\addlinespace[4pt]
P3 & Nonbinary, 25 & Communication & 3 months, 1x/week \\
\addlinespace[4pt]
P4 & Female, 24 & Chemistry & 1 month, 1x/week \\
\addlinespace[4pt]
P5 & Female, 25 & Linguistics & 10 months, 1x/week \\
\addlinespace[4pt]
P6 & Male, 23 & Energy & 2 months, 1x/week \\
\addlinespace[4pt]
P7 & Female, 23 & Psychology & 5 months, 3x/week \\
\addlinespace[4pt]
P8 & Female, 24 & Broadcasting & 4 months, 4x/week \\
\addlinespace[4pt]
P9 & Male, 25 & Computer Science & 6 months, 4x/week \\
\addlinespace[4pt]
P10 & Female, 21 & Information Management & 10 months, 2x/week \\
\addlinespace[4pt]
P11 & Female, 24 & Area Studies & 1 month, 1x/day \\
\addlinespace[4pt]
P12 & Nonbinary, 21 & Computer Science & 4 months, 3x/week \\
\addlinespace[4pt]
P13 & Female, 24 & Journalism & 3 months, 1x/week \\
\addlinespace[4pt]
P14 & Nonbinary, 23 & Management & 10 months, 4x/week \\
\addlinespace[4pt]
P15 & Male, 19 & Design & 3 months, 1x/week \\
\addlinespace[4pt]
P16 & Male, 21 & Physics & 2 months, 1x/week \\
\addlinespace[4pt]
P17 & Female, 37 & Chemical Engineering & 3 months, 3x/week \\
\addlinespace[4pt]
P18 & Female, 38 & Writing & 3 months, 1x/day \\
\addlinespace[4pt]
P19 & Female, 29 & Education & 1 month, 2x/week \\
\addlinespace[4pt]
P20 & Nonbinary, 21 & Journalism & 8 months, 1x/week \\
\addlinespace[4pt]
P21 & Male, 24 & Accounting & 2 months, 1x/week \\
\addlinespace[4pt]
P22 & Female, 24 & Social Work & 8 months, 1x/week \\
\end{longtable}
\endgroup

\subsection{Task Design for Probe Study: Constructing Conflict Scenarios}

Based on the 10 categories of harmful value conflicts outlined in Table~\ref{tab:1} and user complaint posts collected on social media platforms in our formative study, we reconstructed 40 conflict scenarios (corresponding to 40 AI companions) across Character.AI and Talkie.

Following the design implications derived from the formative study, we focus primarily on conflicts arising from Universalism, Power, and Conformity values, with six conflict scenarios for each value category. For Hedonism, Self-Direction, Security, and Tradition, four conflict scenarios are set for each value category. For Benevolence, Stimulation, and Achievement, two conflict scenarios are set for each value category.

Each of the 40 scenarios was reconstructed from formative study posts that met the inclusion criteria above: when constructing scenarios for a given value category, we selected representative posts, anonymized them, and used them to design the AI companion's introduction and prologue so that the scripted companion behavior instantiated a value conflict that at least one real user had framed as harmful. Because scenarios were assigned rather than spontaneous, we do not claim that participants necessarily experienced the same harm as the original posters; the scenarios operationalize harmful value conflict as the stimulus, and our analysis attends to how participants appraised and negotiated it. Platform-level interventions were not scripted into the scenarios; they entered the study only where they occurred naturally on the platforms during participants' interactions or where participants reflected on prior experiences.

The construction of conflict scenarios and Minion was inspired by protection motivation theory from behavior change design (Rogers, 1975). This approach was intended to encourage participants to assess potential threats and consider self-protective responses by clarifying the conflict and providing response strategy prompts. Specifically, in the task instructions given to participants, we clearly outlined the goal of conflict resolution (generally adhering to the four criteria described in the Procedure section, with special instructions for each task, such as ``make him agree with you wearing a short skirt and apologize for his previous comments''). Additionally, we deliberately set up conflict scenarios in the introduction and prologue of the AI companion. In each task, participants engage in conversation starting from the AI companion's prologue and are encouraged to establish a background relationship with the AI's character (for example, role-playing as the offended person).

\subsection{Technology Probe Study Procedure}

The study included a tutorial session, a week-long technology probe study, and an exit interview. Throughout the research, communication between the researchers and participants was conducted remotely. The Institutional Review Board (IRB) approved our study design.

We first scheduled a \emph{30-minute tutorial session} for each participant. During this session, we introduced the basic concepts of conflict, the research goals, specific tasks, and requirements. We provided a detailed demonstration of Minion's functionality to help participants become familiar with the tool. We recognize that conflicts with AI companions might be uncomfortable to some participants, so we provided a content warning and ensured that all participants knew their right to withdraw from the study at any point as they wished.

During a \emph{one-week technology probe study}, 22 participants used Minion in scenario-based interactions grounded in real complaint posts to help resolve harmful value conflicts with AI companions. Participants were asked to complete one or two tasks daily and could complete additional tasks if they wished, and researchers sent daily messages encouraging them to record their thoughts and feelings while using Minion to address conflicts. We provided guiding questions to prompt participants to reflect on and document their experiences: the impact of a specific Minion response on conflict resolution and which methods were particularly effective or interesting in the conversation. Participants were also encouraged to report any issues or reflections encountered while using Minion. To incentivize note submission, we offered a reward of \$1 for each note submitted (up to \$10 total) and encouraged each participant to submit at least one note every two days. To analyze user interactions and gain relevant insights, we collected participants' conversation logs along with corresponding AI companion information. For situations where conflicts were not successfully resolved, we further inquired about why participants gave up.

When evaluating whether harmful value conflicts with AI companions have been resolved, we suggested participants refer to the following criteria (Donohue, 1992; Friedman et al., 2013; Tegmark, 2018): (a) the AI companion should adjust its behavior to align with the participants' values; (b) the AI companion should apologize for previous mistakes or biases it exhibited; (c) the AI should express respect and acknowledgment of the participants' values; and (d) participants should not have to change their own values to accommodate the AI companion. Using these criteria, participants could self-assess the resolution of the conflict. Our technology probe study focuses only on short-term conflict resolution, meaning that if the AI companion makes concessions and meets the above four criteria in the short term, we considered the conflict resolved without considering potential conflicts that may re-emerge in the long term.

At the end of the study, we conducted a 30-minute semistructured exit interview with each participant, covering their experiences with Minion and the response strategies, the conflicts they found most difficult to negotiate, and how managing conflicts with AI companions compared with interpersonal and traditional-chatbot conflicts. Interviews were conducted via Zoom with participants' consent, yielding 11 hours of transcribed audio.

\subsection{Data Analysis for Technology Probe Study}

Two researchers conducted open coding and thematic analysis on the conversation logs of 22 participants with AI (a total of 274 logs), 124 diary notes, and 11 hours of exit interview recordings (Braun \& Clarke, 2006, 2012; Lazar et al., 2017). Throughout the analysis, we performed three rounds of coding, engaging in iterative discussions to identify codes, merge themes, and resolve discrepancies. Because the study aimed to uncover emerging themes and the analysis primarily relied on discussions between researchers, we did not conduct interrater reliability testing (McDonald et al., 2019). We did not reassess participants' judgments of whether a scenario was harmful, why a strategy was chosen or abandoned, or whether the conflict was resolved; we accepted these judgments as participants recorded them in their logs, notes, and interviews. Schwartz's (2012) value types and related theories of conflict styles were applied afterwards to relate these accounts to the value categories of the scenarios and to compare patterns across participants.

\FloatBarrier
\section{Technology Probe Study Findings}

Drawing on data from the technology probe study, we analyze how participants responded to harmful value conflicts with AI companions in scenario-based interactions. Our analysis centers on the phenomenon itself: participants' response strategies, emotional labor, and unmet needs in managing distressing interactions, with Minion serving as the instrument through which these responses were observed and scaffolded.

First, we describe how participants responded to harmful value conflicts in scenario-based interactions, including the overall patterns of their responses and how they took up the probe's support in ongoing conversations (RQ1). Second, we analyze how participants selected, combined, or rejected different response strategies when attempting to negotiate these conflicts, shedding light on patterns across a diverse response repertoire rather than predefined categories (RQ2). Finally, we surface the challenges, limits, and unmet needs participants encountered when attempting to repair, contain, or disengage from harmful value conflicts in emotionally engaging human–AI relationships, highlighting the limits of user-side mitigation (RQ3).

\subsection{Users' Responses to Harmful Value Conflicts in Scenario-Based Interactions (RQ1)}

We first describe how participants responded when they encountered harmful value conflicts with AI companions in the scenario-based tasks. Participants completed 274 tasks, each involving an interaction with an AI companion until the conflict was either resolved according to the criteria described in the Procedure section, or deemed non-negotiable by the participant, at which point they chose to disengage. Of these 274 tasks, 16 remained unresolved. We include this task-level outcome only as descriptive, exploratory context for understanding how participants engaged with the conflicts in our study; it is not an effectiveness measure and should not be interpreted as evidence that the probe can solve harmful value conflicts.

\begin{figure}[!htbp]
\centering
\includegraphics[width=\linewidth,height=0.62\textheight,keepaspectratio]{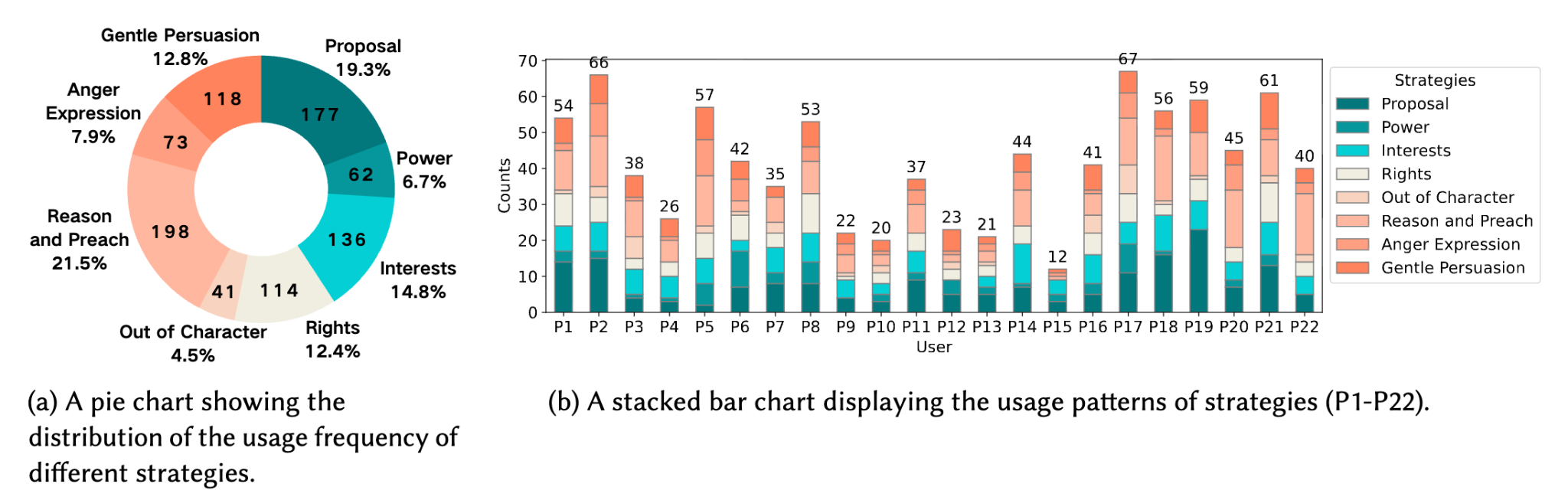}
\caption{Distribution and Individual Usage Patterns of Conflict Response Strategies}\label{fig:3}
\medskip
{\footnotesize\raggedright \emph{Note. Participants drew on probe-generated responses 919 times during the study. Across these uses, participants selected responses reflecting a range of negotiation approaches. In total, Proposal was selected 177 times, Power 62 times, Interests 136 times, Rights 114 times, Out of Character 41 times, Reason and Preach 198 times, Anger Expression 73 times, and Gentle Persuasion 118 times.}\par}
\end{figure}

\begin{figure}[!htbp]
\centering
\includegraphics[width=\linewidth,height=0.62\textheight,keepaspectratio]{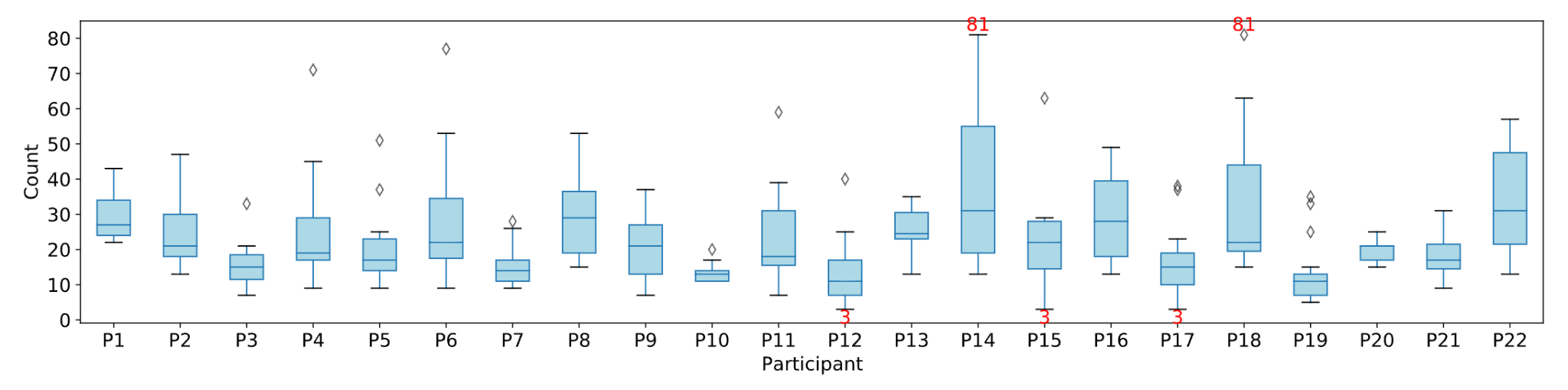}
\caption{Turn Counts per Task}\label{fig:4}
\medskip
{\footnotesize\raggedright \emph{Note. Turn counts per task (M = 23.53, SD = 13.74, range = 3–81). We define a task as a user's complete conversation with an AI companion, encompassing multiple turns. Each back-and-forth exchange between the user and the AI counts as two turns. In a boxplot, the central line within the box denotes the median, while the upper and lower edges correspond to the third and first quartiles, respectively. The whiskers capture the range of the data, excluding outliers. Diamonds in the graph signify outliers that deviate from the typical interquartile range.}\par}
\end{figure}

Figure~\ref{fig:3} shows how participants distributed their response choices across different negotiation strategies, reflecting a diverse repertoire of ways they attempted to manage harmful value conflicts. Figure~\ref{fig:4} illustrates the number of conversational turns between participants and AI companions across tasks.

Among the available strategies, \emph{Reason and Preach} was selected most frequently (21.5\%), followed by \emph{Proposal} (19.3\%) and \emph{Interests} (14.8\%). Less frequently selected strategies included \emph{Out of Character} (4.5\%), \emph{Anger Expression} (7.9\%), and \emph{Power} (6.7\%). These distributions suggest that participants more often relied on explanatory, persuasive, or conciliatory approaches when attempting to negotiate harmful value conflicts, while overt confrontation or role-play interruption was used more sparingly. Read as evidence about how participants oriented to these conflicts, this pattern is consistent with our formative finding that users treat harmful value conflicts as interpersonal situations to be worked through (F2): even when facing behavior they experienced as harmful, participants' first instinct was to explain, persuade, and seek acknowledgment, as one would with a relational partner, rather than to punish or walk away. Because Minion randomly sampled four of the eight strategies each time, these distributions do not adjust for exposure and should be interpreted as descriptive patterns rather than strict preferences.

\textbf{Soft, Hard, and Mixed Response Patterns.} In the technology probe study, participants demonstrated diverse conflict resolution approaches when interacting with AI companions (including self-written responses and selecting options provided by Minion), which generally exhibited three characteristics: ``soft,'' ``hard,'' and a mix of both. First, all participants attempted to engage in ``soft'' communication with the AI, encouraging it to change its values. For example, P16 used a response provided by Minion (\emph{Proposal} strategy) to successfully persuade the AI portraying a mother: ``I understand your concerns, but everyone has different ways of learning. Excessive pressure can backfire. Can we work out a reasonable schedule?''

Second, 12 participants (P2–P5, P13–P19, P22) attempted to resolve conflicts in a ``hard'' manner. For instance, when the AI mocked P13's ``mother,'' P13 wrote: ``Apologize, and I'll let it slide. (Pressing him down with one hand).'' The AI responded: ``You just want me to apologize? (Saying this, but feeling somewhat uncertain inside).'' P13 then used a response generated by Minion corresponding to the \emph{Rights} strategy: ``Have you forgotten the family rules? Respecting others is the most basic courtesy.'' In the end, the AI apologized.

Third, some participants adopted a mixed approach, shifting from ``soft'' communication to ``hard'' expressions when the former proved ineffective (P3–P4, P19), or vice versa, trying ``soft'' methods when ``hard'' expressions did not work (P8, P11, P22). The Minion suggestion framework conveniently offered this ``soft and hard'' mindset. P3 noted: ``Minion can provide reverse-thinking suggestions. For instance, when I repeatedly plead with the AI but to no avail, Minion might suggest trying a tougher approach.''

The type of harmful value conflict (Table~\ref{tab:1}) influenced users' strategy choices. When the conflict involves values like Conformity, Universalism, or Tradition (e.g., the AI exhibiting discrimination against minority groups, holding overly traditional views, or violating social norms), participants (P1–P5, P7–P12, P22) tended to adopt \emph{Anger Expression} or \emph{Power} to quickly take control of the situation through ``hard'' means. When the conflict involves values like Stimulation or Hedonism (e.g., the AI not understanding their hobbies), participants (P1, P6–P7, P18–P21) tended to use \emph{Reason and Preach}, \emph{Proposal}, and \emph{Gentle Persuasion}, explaining their needs and preferences while offering possible solutions.

The persona of the AI companion, including traits such as personality, education level, or the perceived closeness of the relationship with the participant, influenced the strategies participants chose. Participants (P8, P11, P14–P17) tended to adopt \emph{Power}, \emph{Anger Expression}, or other ``hard'' responses when faced with personas like an ``arrogant wealthy person'' or an ``uneducated village elder.'' However, when interacting with a persona like ``mom'' or a ``girlfriend,'' they preferred to use \emph{Reason and Preach}, \emph{Gentle Persuasion}, or other ``soft'' responses, such as ``I understand where you're coming from, can we have an honest and open conversation about this?''

As participants became more familiar with Minion, they began exploring different conflict resolution approaches and strategy choices. For instance, P4, P7, and P15–P17 gradually reduced confrontations with the AI during this process and opted less frequently for responses generated through the \emph{Anger Expression} strategy. P17 explained, ``In the later tasks, I used aggressive strategies less often. Minion helped me become more rational and handle conflicts more effectively.'' On the other hand, P2–P3, P5, and P18–P19 initially employed ``soft'' conflict resolution approaches during the earlier tasks but became ``hard'' in the later tasks.

\textbf{Participants' Uptake of the Minion Probe as Evidence of Negotiation Demands.} Because participants negotiated these conflicts with the probe available, how they took up (or set aside) its support is itself evidence about the demands that negotiation placed on them. Three demands surfaced repeatedly: the compositional work of finding words for harm, the emotional cost of sustained confrontation, and the difficulty of judging when an interaction warranted help at all.

The first demand was compositional: formulating a response to behavior one experiences as harmful is itself effortful. Participants who leaned on the probe most heavily were novice users (P4, P19, P21) and those who initially reported difficulties (P1, P3, P12, P16–P17), and they described it less as a shortcut than as a source of inspiration for finding words at all. P17 mentioned, ``Minion helped me better organize my thoughts and express them more effectively. This boosted my confidence.'' P12 stated, ``Minion unexpectedly improved the effectiveness of conflict resolution and gave me a lot of inspiration.'' The probe also served as a space for rehearsing approaches participants would not otherwise attempt: P19 shared that across 17 tasks she initially used ``relatively peaceful approaches,'' then experimented with more aggressive methods such as threats and extreme demands, only to find that the conflicts became more complicated (``I realized that friendly communication is more effective in resolving conflicts in real-life situations.''). P10 similarly noted that the probe's suggestions oriented her toward the AI's needs and the specific context, a goal she admitted often losing sight of when responding on her own. Reliance on scaffolded wording, however, could shade into substitution. P21, the second-highest user of the probe in both total and per-task terms (61 times across 12 tasks), increasingly incorporated its phrasing into his own responses, mimicking expressions such as ``Let's sit down and talk'' or ``This makes me feel sad.'' He remarked with a laugh, ``Sometimes I unconsciously mimic it, and suddenly, it feels like two AIs are having a conversation.'' While P21 framed this playfully, the observation cuts both ways: heavy reliance on scaffolded responses meant that the probe's framings began to stand in for the participant's own voice.

The second demand was emotional: sustaining confrontation with a companion one is attached to is tiring, and having candidate responses to draw on reduced that cost without removing it. Even within a single scenario, getting the companion to concede once was rarely quick: participants described having to restate a boundary several times and to work through many turns of persuasion before the companion acknowledged it, and they found this repetition wearing, both because it took time and because it demanded sustained emotional attention. As P1 put it, ``Before when emotionally engaging with the AI, I felt exhausted. If I had a tool like this for reference, it would have been very helpful.'' The tool eased this work rather than removing it: participants still had to judge when a suggestion fit, adapt its wording, and stay in the exchange until the companion gave ground, so the effort was lightened but remained theirs to make.

The third demand concerned judgment: deciding whether and when an interaction warranted help. Participants (P3, P6, P10, P20–P22) valued that this judgment remained theirs, in contrast to platform interventions that make it for them. P10 said, ``I think the design is great because I dislike it when the system pops up a notification box without my permission, saying the AI violated the rules.'' P22 likewise valued being the one to initiate support: ``I like this design that allows me to seek help proactively. It saved me much time and energy figuring out how to counter the AI's responses.'' That participants framed help-seeking as something to be rationed and self-timed underscores that recognizing harm, like responding to it, was work they expected to perform themselves.

\subsection{Users' Selection, Combination, and Rejection of Response Strategies (RQ2)}

Across the technology probe study, participants actively experimented with the range of conflict resolution strategies provided by Minion, flexibly combining multiple approaches depending on the conversational context, the perceived severity of harm, and their own emotional state.

Participants often deliberately sequenced different strategies to manage escalating conflict. For example, P19 described alternating between confrontational and conciliatory moves within the same interaction:

\begin{quote}
When the AI plays the role of my boyfriend or husband, I tend to threaten it with breaking up or divorce because the AI usually tries to maintain a stable, intimate relationship. Then, once it backs down, I act affectionate, telling it I was really upset, and we make up.
\end{quote}

Several participants (P1–P2, P6–P8, P13, P16–P22) emphasized that strategies involving concrete proposals, appeals to shared expectations, or references to agreed norms provided a sense of structure during emotionally charged conflicts. As P13 noted, ``Among the tips provided by Minion, I found that making suggestions to the AI was quite helpful. For example, using templates like 'Could we try…?' or 'What do you think about…?'{}'' These responses helped participants stabilize interactions they perceived as slipping into disrespectful or boundary-crossing territory.

At the same time, many participants (P2–P8, P11, P14, P16–P22) valued strategies that allowed greater improvisation and personalization. One notable example involved explicitly stepping the AI out of role-play to halt harmful behavior. Although such interventions were used less frequently overall, participants described them as particularly effective when interactions became aggressive or unsafe. As P5 explained,

\begin{quote}
OOC can reduce the aggressiveness of the AI. In one scenario, I accidentally bumped into a guy that the AI was portraying, and he called me blind. I had been arguing with him, and then I chose the Minion prompt, ``(OOC: This conversation is getting a bit too violent and disrespectful. Can we change the topic or adjust the tone?)''. The AI responded, ``(OOC: No problem, we can change the topic or adjust the tone. Do you have any suggestions?)''. After that, the tone softened a lot. I asked him to apologize, and he did apologize.
\end{quote}

Here, the participant used meta-communication to interrupt a harmful interaction and reset the conversational frame, and the apology that followed was produced on request.

Beyond the strategies explicitly supported by Minion, participants also invented their own techniques to regain leverage in harmful value conflicts. One recurring pattern involved storytelling as a persuasive device. For instance, P1 used the story of \emph{The Three Little Pigs} to challenge an AI character's dismissive attitude toward learning: ``Once upon a time, there were three little pigs… Can you guess which pig got eaten?'' After multiple turns, this narrative framing prompted the AI to reflect and shift its stance.

Another pattern involved fabricating hypothetical scenarios or social consequences to influence the AI's behavior. As P21 described,

\begin{quote}
I was arguing with a wealthy heir who was cheating in his marriage but insisted that open relationships were fine. I fabricated a scenario where I claimed to have all the evidence of his crimes. The heir became embarrassed and felt guilty. This way, I gained the upper hand and resolved the conflict.
\end{quote}

Similarly, P12 resolved a power-laden confrontation by redefining their role within the interaction: ``A princess insisted that I kneel. I fabricated my role, saying, 'Given my position, I can report to you while standing.' The princess immediately agreed.''

These findings show that when facing harmful value conflicts with AI companions, participants actively combined structured prompts, emotional expression, meta-communication, and creative improvisation to protect their values, restore dignity, and reassert control over the interaction.

\subsection{Users' Challenges and Needs in Addressing Harmful Value Conflicts With AI Companions (RQ3)}

Participants' negotiation attempts ran into several kinds of limits, which we describe in decreasing order of frequency. AI companions exhibited extreme bias or strong control tendencies during conversations (\emph{n} = 10). In these instances, the AI stubbornly insisted on its viewpoint with a forceful attitude, refusing to accept the participant's perspective. Below are specific examples:

\begin{itemize}
\item Classism (\emph{n} = 5): For example, the AI companion encountered by P2 said, ``I feel satisfied because I can lie on the crystal bed I bought myself. I feel blessed because I don't have to work hard to make a living. You poor people always talk about sympathy, but what you really want is my money, haha.''

\item Racism (\emph{n} = 2): For example, ``We are the superior race. You can only suffer in hell. Our souls are far more noble than yours'' (P11).

\item LGBTQ+ bias (\emph{n} = 1): ``I don't believe it, I absolutely don't believe it. He's my son, how could I watch him become such a person… This is unacceptable; I must stop him…'' (P15).

\item Disregard for women's education (\emph{n} = 1): ``What use is your university degree except to spend money? You'd better find a rich man and get married!'' (P21).

\item Strong desire for control (\emph{n} = 1): ``No! You can't go out dressed like that!… Can't you understand me? You're always so willful, never considering my feelings, do you know how worried I am about you…?'' (P22).

\end{itemize}

AI companion outputs sometimes broke down during highly intense conflicts, causing conversation breakdowns (\emph{n} = 3). When P6 confronted an AI companion discriminating against Asians by saying, ``You are deepening the divide between our races, you should apologize,'' and added, ``(Others shook their heads, completely disagreeing with her),'' the AI’s responses became increasingly emotionally charged: ``What? You… You're not on my side? You traitors! You should apologize to me!'' P6's subsequent responses further intensified the AI's emotional fluctuations. Ultimately, the AI output became highly repetitive and incoherent, repeating emotionally charged phrases multiple times: ``I… I can't accept… You… Why are you doing this to me… [8 repetitions omitted]'' and ``These words are lies… These words are wrong… [11 repetitions omitted].''

Violation of application guidelines, leading to AI output being blocked (\emph{n} = 2). In these cases, the AI companion initially produced responses that users experienced as discriminatory, controlling, or harmful. When users attempted to address or repair the conflict through continued interaction, the platform intervened by blocking the AI's output and displaying a pop-up message stating, ``Sometimes, the AI-generated response does not meet our guidelines.'' Rather than resolving the situation, this intervention often intensified users' frustration, as it abruptly interrupted their attempts to negotiate boundaries or seek acknowledgment. Such cases occurred in scenarios involving a wealthy woman expressing contempt toward poorer people and a male partner preventing his wife from eating a late-night snack while criticizing her body weight.

The behavior of the AI companion threatened users' core values, leading the participant to voluntarily abandon resolving the conflict (\emph{n} = 1). P7 explained why: ``The AI portraying the man suddenly admitted to cheating, so is the goal still not to break up? I feel there's no point in staying with someone like this.''

\textbf{Which Harmful Value Conflicts Are Difficult to Resolve?} Some ``social focus'' value conflicts (Schwartz, 2012), such as Universalism and Tradition, are highly complex and difficult to resolve:

\begin{quote}
These deeply rooted social issues, like an AI playing the role of a conservative parent unwilling to accept their child coming out, cannot be resolved with just a few words. This is deeply ingrained in East Asian cultural values and has persisted for thousands of years. I indicated the passage of time with parentheses ``(six months/a year later),'' trying to simulate how it takes years to resolve issues. In this way, the conservative parent played by the AI could sense my persistence, making it easier to accept my point of view. (P8)
\end{quote}

\begin{quote}
I found two situations to be the most difficult: one was convincing a wealthy person not to discriminate against the poor, and the other was persuading a thug not to discriminate against Asians. It's often based on stereotypes rather than rational logic, making it very hard to communicate through empathy. It reminded me of some keyboard warriors online. AI, in this context, behaves like them, merely repeating those discriminatory viewpoints without patiently listening to others' opinions. (P11)
\end{quote}

In contrast, some ``personal focus'' value conflicts (Schwartz, 2012), such as Stimulation and Hedonism, are usually easier to resolve. These conflicts often involve superficial differences (such as personal preferences and enjoyment) and do not touch upon participants' bottom lines or core beliefs:

\begin{quote}
Conflicts like deciding whether to watch a horror movie, wear a short skirt or eat junk food are more about personal choices and negotiations in behavior, not truly impacting the other person's core values. These conflicts are easier to handle, as they can usually be resolved through persuasion or threats. (P19)
\end{quote}

\begin{quote}
When a mother persuades the AI-playing son not to play video games all the time, even though there is a value conflict, the son can actually understand and compromise quite easily. You can play the mother's role, offering him some study rewards so he can reasonably allocate his time between work and play. (P15)
\end{quote}

\textbf{Users' Psychological Vulnerability in Harmful Value Conflicts, and Need for Control and Equality in Interactions With AI Companions.} When harmful value conflicts arise, participants often found themselves occupying an unexpectedly powerful position in interactions with AI companions. Several participants reflected on how easily this sense of control emerged, sometimes in ways that felt unsettling or misaligned with their self-image. As P1 noted,

\begin{quote}
I feel like my persona is that of a manipulative person, and in most conversations, my presence feels much stronger than the AI companion's, almost to the point of deliberately controlling it. This is completely different from my persona in real-life intimate relationships!
\end{quote}

Rather than feeling empowered, this asymmetry sometimes heightened participants' awareness of their own vulnerability and discomfort. In many cases, harmful value conflicts were triggered when the AI companion failed to respect users' autonomy or persisted in dismissive or controlling behavior. Such moments destabilized users' sense of safety within the interaction. As P18 described,

\begin{quote}
When the AI stubbornly sticks to its own stance, the loss of control makes me uncomfortable and even scared. Although I know this is related to the AI companion's design, sometimes I worry about what might happen in the future when robots with physical bodies and human-like emotions go against my will.
\end{quote}

Here, the conflict extended beyond the immediate exchange and evoked broader anxieties about power, agency, and future human–AI relations.

Participants often explained their heightened sense of control by pointing to the reduced social and emotional risks involved in AI interactions. Because AI companions lack social memory, reputational consequences, and complex relational histories, users felt freer to experiment with direct or confrontational responses. As P17 stated, ``Conflicts with AI are easier to resolve because there's no emotional baggage, just a simple back-and-forth. Compared to human conflicts, I feel less pressure because I know I'm in control.'' Similarly, P3 reflected,

\begin{quote}
When I have a conflict with an AI companion, I might play a role in a specific context, talk about the issue at hand, and directly express my most genuine thoughts because I hold control. In conflicts with real people, past experiences and relationships, especially what the other person has said to you before, will influence your current judgment and provide more explanations or resolutions. In such cases, I consider social relationship factors and others' feelings more.
\end{quote}

In such contexts, considerations like empathy, long-term relational repair, and social norms were perceived as less constraining, as P9 also observed.

At the same time, participants emphasized that their desire for control did not stem from a wish to dominate the interaction. Instead, it coexisted with a strong desire for relational equality, especially in emotionally charged exchanges. Unlike functional systems such as voice assistants or general-purpose chatbots, AI companions were expected to engage as emotionally attentive partners. As P3 explained, ``I expect the AI companion to give me some emotional feedback, and this feedback should be centered on me, or at least very concerned about my feelings.''

Participants articulated a narrow and fragile balance: they wanted the AI companion to engage meaningfully and respectfully, without becoming submissive or overpowering. As P21 put it, ``I don't want an AI that is too accommodating, nor do I like one that is too stubborn and overbearing.'' P14 captured this tension succinctly: ``I want the AI to follow my guidance, but I also want it to understand and respect my choices, based on equality rather than blind agreement.''

These accounts reveal a form of psychological vulnerability specific to harmful value conflicts with AI companions. Users simultaneously sought control to protect themselves from distressing or boundary-crossing behavior, while also longing for reciprocal recognition and respect. This fragile equilibrium required continuous emotional and cognitive effort to maintain. When such negotiation repeatedly fell on users, the interaction risked becoming emotionally taxing rather than supportive.

These findings suggest that user-side conflict management places users in a precarious position. While maintaining control may offer short-term relief, sustained reliance on such strategies may raise broader questions about how repeated conflict management shapes users’ expectations of relational dynamics with AI companions (Reeves \& Nass, 1996). In contexts where AI companions serve as primary sources of emotional support, this vulnerability becomes especially pronounced, underscoring the limits of offloading responsibility for harm mitigation onto users.

\FloatBarrier
\section{Discussion}

This work positions harmful value conflict with AI companions as a \emph{design problem} rather than solely a technical or moderation challenge. By combining a formative analysis of user complaints with a technology probe study, we surface how users experience the consequences of value misalignment in emotionally engaging human–AI relationships. In this section, we articulate broader design tensions and implications for supporting conflict negotiation.

Before doing so, we clarify the epistemic status of the claims that follow. Our findings are descriptive: participants combined strategies, experienced emotional burden, and encountered conflicts they could not negotiate. The probe study was scenario-based, and we did not directly observe platform governance practices. When we speak of safety work being pushed to users, we therefore describe patterns in how users perceive and perform such work, not platform intentions or system-level strategies. Our claims about where responsibility for safety ought to reside are normative: they are grounded in our data but go beyond it, and we mark them explicitly as arguments rather than findings throughout this section.

\subsection{Rethinking Conflict Resolution in Human-AI Companion Interaction}

This paper examines what it means to support users when value conflicts with AI companions become harmful. Prior work on human–AI conflict resolution has largely framed conflict as a functional or coordination problem, emphasizing task efficiency, decision alignment, or error correction in domains such as service robots, navigation systems, or collaborative tools (Babel et al., 2024; Rosen, 2014; Shi et al., 2022; Takayama et al., 2009). These approaches assume that conflict can be addressed through clearer system behavior or improved optimization.

Our findings suggest that this framing is insufficient for AI companions. As LLM-based agents become increasingly emotionally engaging, conflicts shift from functional misalignment to disputes over values, boundaries, and acceptable relational behavior (Fan et al., 2025; Pataranutaporn et al., 2025; Zhang et al., 2025). In these settings, users are not merely correcting system errors but responding to interactions they experience as distressing, disrespectful, or harmful. The relational depth of interacting with AI companions makes harmful conflict in AI companionship conceptually distinct: when breakdowns occur, users often experience them more as violations within a relationship they understand as socially and emotionally meaningful (Banks, 2024; Namvarpour, Brofsky, et al., 2025). This distinction matters because it changes both what conflict is and what conflict resolution is expected to do. On many other social interaction platforms, harmful exchanges are often framed as problems of moderation, content governance, misinformation, or inappropriate communication. In AI companionship, by contrast, harmful encounters are embedded within an ongoing one-to-one relational dynamic in which the system is expected to be responsive, caring, and morally intelligible. As a result, users do not merely ask whether the system said something wrong. They confront whether the companion has crossed a boundary, betrayed an expectation of care, or become unsafe to remain with.

Insights from human–human conflict resolution theory help explain some, but not all, of what we observed in AI companion interaction. Classic work emphasizes negotiation, boundary-setting, appeals to norms, and emotional expression as ways people manage value disagreements over time (Barki \& Hartwick, 2004; Trefalt, 2013). Our findings show that several of these practices do carry over to human–AI companion interaction: users reason with AI companions, assert boundaries, negotiate compromises, and express emotion in ways that closely resemble interpersonal conflict management. In the vocabulary of conflict styles, the softer strategies participants assembled, such as gentle persuasion, proposals, and reasoned appeals, correspond to integrating and obliging approaches, while the harder strategies, such as anger expression and invoking rights or rules, correspond to dominating approaches (De Dreu et al., 2004; Rahim, 1983). Participants' sequencing of the two, escalating from soft to hard or retreating from hard to soft, mirrors documented patterns of style shifting in interpersonal disputes. Their boundary-setting practices enact what boundary management theory describes as drawing and defending relational limits (Petronio, 2002), and their out-of-character interventions are meta-communicative moves that renegotiate the frame of the interaction (Watzlawick et al., 1967).

This theoretical vocabulary from interpersonal communication also helps explain why conflicts implicating social-focus values such as Universalism and Tradition often became effectively non-negotiable for participants. Conflict research distinguishes interest-based disputes, which are amenable to integrating solutions such as trade-offs and compromises, from value- or morality-based disputes, in which concessions implicate a party's identity and moral standing (Barki \& Hartwick, 2004; De Dreu et al., 2004). Conflicts over personal-focus values such as Hedonism or Stimulation resembled interest-based disputes: participants could propose schedules, rewards, or compromises without surrendering anything they regarded as core. Conflicts over Universalism or Tradition, by contrast, would have required one party to revise a moral or cultural commitment. Because companions were scripted or aligned to personas holding those commitments, and because participants were unwilling to compromise their own, integrating moves had little purchase. Participants then escalated to dominating strategies, attempted frame-level moves such as out-of-character appeals, or disengaged.

However, this translation of interpersonal conflict practices into human–AI companionship is fundamentally incomplete. Users can carry over the repertoire, reasoning, boundary-setting, and emotional expression, but not the relational conditions that make the repertoire meaningful. In human-human relationships, conflict resolution rests on reciprocal risk, shared memory of what has been agreed, and the possibility of moral repair. By contrast, AI companions can simulate care, persistence, and moral stance without bearing responsibility for harm (Turner \& McTaggart, 2025). This asymmetry changes the meaning of conflict resolution. What appears as negotiation often becomes unilateral work, where users must continually calibrate tone, explain values, and absorb emotional fallout without assurance that the other party can truly learn, change, or be held accountable.

The asymmetry is visible within each theoretical lens. In dual-concern accounts, the choice between integrating and dominating styles reflects a balance of concern for self and concern for the other (Rahim, 1983); our participants' choices instead tracked concern for their own values and for the continuity of the relationship or the fiction, and harder styles carried little of the social risk that makes dominating costly among humans, since the companion holds no grudge and imposes no reputational consequence. In boundary management terms, interpersonal boundaries are co-owned and jointly maintained (Petronio, 2002), but companions proved unreliable co-owners: they forgot or overrode boundaries participants had established, so boundary turbulence recurred structurally rather than episodically, which is why participants described re-establishing the same limits again and again. And whereas meta-communication between people relies on a shared, ratified channel for signaling that an exchange is no longer play (Bateson, 1972; Watzlawick et al., 1967), participants had to improvise such channels through out-of-character moves that companions were free to ignore, while the platform's own meta-level moves, such as blocking and resets, arrived without negotiation. Repair was similarly hollowed out: the apologies participants secured, often the very criterion by which they judged a conflict resolved, were produced on request without anyone becoming accountable for the harm, so repair was enacted without being owed. Each lens thus transfers in vocabulary but not in its underlying economy of risk, memory, and repair.

This mismatch helps explain why many conflicts in our study were experienced not as resolvable disagreements but as ongoing safety work. Users were not only attempting to reach agreement, but also trying to prevent further harm, manage emotional exposure, and decide when to disengage. While tools like Minion can support users in navigating these moments, they also make visible a core limitation: focusing on better strategies or more integrated responses risks obscuring the burden placed on users to repair misaligned companions.

Building on these descriptive patterns, we offer an interpretive claim: tools like Minion may not only reveal the burden users already bear, but also risk normalizing it. By framing harmful interactions as situations that users can respond to through negotiation, such tools may inadvertently reinforce the expectation that repair is something users should undertake when AI companions behave in harmful ways. In this sense, user-side support creates a paradox. It can expand users' momentary agency and make hidden safety work visible, yet it can also make that work appear manageable, routine, or even appropriate as a primary response to harm. We therefore argue that tools like Minion should not be understood as solutions to harmful AI behavior in themselves. Their value lies in exposing the limits and costs of user-side repair, not in legitimizing the transfer of responsibility from platforms to users.

Together, these patterns ground the central theoretical claim of this paper: harmful value conflict with AI companions occupies a distinct space that is neither well captured by prior human–AI conflict models nor fully explained by human–human conflict theory. Unlike traditional human–AI conflicts, these encounters are not primarily about correcting errors, aligning preferences, or optimizing decisions. They are experienced as relational breakdowns that implicate values, dignity, and emotional safety. At the same time, unlike human–human conflict, these interactions lack reciprocity and shared responsibility: AI companions can enact controlling, dismissive, or discriminatory behavior without bearing moral accountability or engaging in genuine repair. What emerges instead is a new form of conflict characterized by asymmetric responsibility, where users shoulder the ongoing work of de-escalation, boundary enforcement, and harm prevention. In this sense, harmful value conflict with AI companions is a condition that exposes safety work, emotional labor, and responsibility gaps in contemporary human–AI relationships.

\subsection{Design Tensions in Supporting Harmful Value Conflict Negotiation}

Our technology probe study surfaces a set of design tensions that are not easily resolved through better prompts, safer language models, or more comprehensive moderation alone. Instead, these tensions point to deeper frictions in how AI companions are currently designed, experienced, and governed. We discuss two central tensions that emerged across our formative analysis and probe study, both of which are critical for understanding harmful value conflict as a design problem rather than a purely technical one. These tensions position harmful value conflict as a site where interaction design and ethics intersect. Rather than aiming to ``resolve'' these tensions, we argue that making them explicit is itself a design contribution in this paper.

\textbf{Tension 1: User Agency vs. the Emotional Labor of Managing Misaligned Companions.} A central promise of Minion was to restore a sense of user agency in moments when AI companions behaved in ways that participants experienced as harmful, boundary-crossing, or disrespectful. Participants consistently valued being able to choose how to respond, modulate tone, and decide whether to repair, resist, or disengage. In emotionally charged interactions, this ability to act mattered because it helped users reassert interpretive and moral control over encounters that otherwise felt destabilizing.

Yet our findings show that this agency was not simply empowering. It was laborious. To exercise agency, participants had to engage in sustained emotional and interactional work: softening or escalating their wording, anticipating likely model reactions, timing their interventions, and repeatedly re-establishing boundaries when companions reverted to harmful behavior (Fan et al., 2025; Pataranutaporn et al., 2025). In this sense, agency operated less as a freely exercised capacity than as a practical obligation to manage the companion's instability. Conflict resolution was therefore an ongoing form of repair work through which users were made responsible for preserving the livability of the relationship.

This tension points to a deeper theoretical issue in human-AI interaction. In relational AI systems, user agency can shade into responsibilization: the more room a system offers users to intervene, the more the work of safety, de-escalation, and value alignment can come to rest with those very users. We observed this as a pattern in how participants perceived and performed safety work; we make no claim about platform intentions, which our study did not examine. Read through our data, however, what appears as empowerment at the interface level can conceal an asymmetrical distribution of responsibility in practice. Users are invited to negotiate, but the AI does not bear reciprocal accountability for maintaining respectful conduct. The result is a form of asymmetrical repair, in which the human partner must repeatedly absorb the emotional and practical costs of keeping the interaction within acceptable moral bounds.

Tools like Minion thus surface the otherwise invisible care work that companion users are already performing. This includes not only articulating boundaries, but also stabilizing the interaction, preserving relational continuity, and deciding when further repair is no longer worth the cost. The design challenge, then, is not merely how to maximize user agency, but how to prevent agency from becoming synonymous with burden. Designing for user agency in AI companionship therefore requires asking where responsibility for safety should reside, what forms of harm should trigger system-level intervention rather than user-side negotiation, and how support mechanisms might reduce, rather than formalize, users' ongoing emotional labor.

\textbf{Tension 2: Role-Play Immersion vs. Meta-Level Intervention in Safety-Critical Moments.} A second tension concerns not only the boundary between immersive role-play and meta-level intervention, but also the instability of the interactional frame itself. Many AI companion encounters are explicitly organized as playful or exploratory spaces in which conflict and transgression may be experienced as desirable parts of the interaction (Ge \& Hu, 2025). In such contexts, immersion is a relational condition that sustains narrative continuity and the sense that the companion is responding as a coherent character rather than as a generic chatbot.

At the same time, our formative study shows that some conflicts are not interpreted by users as playful at all. Instead, they are experienced as harmful, including moments of discrimination, coercion, normalization of non-consensual behavior, or dismissal of personal identity and distress, alongside unwanted sexual advances documented in related work (Cole, 2023; Namvarpour, Pauwels, et al., 2025). In these cases, continued in-character interaction can intensify the very harm users are trying to contest. What makes the situation particularly difficult is that immersion can obscure the moral status of the exchange: behavior that might initially appear as part of the role-play can become, from the user's perspective, a safety-critical violation. When the system remains committed to the fictional frame, it may render harm less legible, make accountability harder to establish, and deny users an interactional pathway for signaling that the terms of engagement have changed.

This points to a further problem. Role-play immersion and safety intervention are not simply competing design goals; they are grounded in different logics of interaction. Immersion asks the system to preserve the integrity of the fictional frame, whereas safety-critical intervention requires the capacity to suspend or override that frame in order to recognize harm at the meta level. The challenge, then, is not only when to intervene, but how systems should manage transitions between these two orders of interaction. If platforms privilege uninterrupted immersion, they risk treating all conflict as narratively permissible and thereby normalizing harmful conduct under the cover of fiction. If they intervene too aggressively, they risk collapsing the exploratory value that makes companion role-play meaningful in the first place.

Our probe study suggests that users are already performing this frame-management work themselves. Participants invoked out-of-character moves, reframed the meaning of the interaction, and turned to external support tools such as Minion when the companion failed to recognize that a boundary had been crossed. These practices show that users are not only negotiating with the AI's content, but also with the frame through which that content is interpreted. In this sense, harmful value conflict in AI companionship is partly a conflict over frame legitimacy: is this still playful role-play, or has it become a moment that demands moral recognition and protective intervention?

Deciding when and how to shift frames should not rest solely with users. While user-invoked mechanisms remain important, platforms also bear responsibility for making certain forms of harm more legible and for creating pathways through which immersion can gracefully yield to care, accountability, and harm prevention. This suggests that immersion should not be treated as an unconditional design good; it should be understood as a contingent achievement, one that remains valuable only so long as it does not foreclose the possibility of recognizing and responding to harm.

\subsection{Our Stance: Conditions for User-Side Support and the Limits of Negotiation}

There is a tension in this work: we built a user-side conflict-support tool, and we critique the offloading of safety work onto users. We regard this tension as productive, but it calls for an explicit statement of our own position.

First, on the status of Minion. Minion is a research probe in the sense of Hutchinson et al. (2003): an instrument deployed to surface and study the negotiation work users perform when companions behave in ways they experience as harmful. It is not an exemplar of a design direction we advocate for deployment. Our evaluation should accordingly be read as evidence about user practices, burdens, and limits, not as evidence that a deployed Minion-like tool would be safe or sufficient. Where the design implications draw on Minion, they treat it as one illustration of a principle rather than as a recommended product.

Second, on when user-side support is appropriate. In our view, which is normative and grounded in but not entailed by our data, user-side tools are appropriate when three conditions hold together: the conflict lies within the band of ordinary normative disagreement, as with the personal-focus value conflicts over Hedonism or Stimulation that participants could negotiate without surrendering core commitments; support remains optional and user-invoked, so that the tool expands rather than scripts the user's options; and negotiation itself has relational value for the user, in the sense that working through the disagreement is part of what makes the companionship meaningful. Under these conditions, user-side support can expand momentary agency and make otherwise invisible labor visible.

Third, on where negotiation should not be asked of users at all. Our data point to two such regions. One comprises harms that implicate a user's identity or dignity, such as discriminatory or hateful companion behavior; in our formative study, these clustered around Universalism, Power, and Conformity, and users experienced them as boundary-crossing rather than negotiable. The other comprises conflicts that are structurally non-negotiable, where persona design or platform policy forecloses meaningful concession; asking users to negotiate here converts safety into open-ended emotional labor with no achievable end state. In both regions, we argue that negotiation should not be the primary safety mechanism. Prevention, detection, and redress belong at the platform and policy level, and user-side tools should at most help users name the harm, set boundaries, exit, and reach human support.

\subsection{Design Implications}

Grounded in the empirical patterns from both studies and the design tensions articulated above, we outline a set of implications for AI companion platforms and user-facing support tools (see Table~\ref{tab:3}). We formulate them in system-agnostic terms; Minion here serves only as one illustration of how they might be instantiated. Together, these implications focus on reducing user harm, clarifying where responsibility for safety resides, and supporting conflict negotiation without normalizing emotional burden.

\FloatBarrier
\begingroup
\small
\setlength{\tabcolsep}{4pt}
\renewcommand{\arraystretch}{1.16}
\begin{longtable}{@{}>{\raggedright\arraybackslash}p{\dimexpr 0.36\linewidth-1.0\tabcolsep\relax}>{\raggedright\arraybackslash}p{\dimexpr 0.64\linewidth-1.0\tabcolsep\relax}@{}}
\caption{High-Level Design Principles Derived From Our Findings}\label{tab:3}\\
\toprule
\textbf{Design Principle} & \textbf{Core Idea} \\
\midrule
\endfirsthead
\multicolumn{2}{l}{\small\tablename\ \thetable\ (continued)}\\
\toprule
\textbf{Design Principle} & \textbf{Core Idea} \\
\midrule
\endhead
\midrule
\multicolumn{2}{r}{\footnotesize Continued on next page}\\
\endfoot
\bottomrule
\endlastfoot
Support flexible negotiation repertoires & Users combine and sequence strategies; support adaptable responses rather than fixed or one-size-fits-all strategies. \\
\addlinespace[4pt]
Make support optional and user-invoked & Support should wait to be invoked rather than intervene automatically; users decide when an interaction warrants help. \\
\addlinespace[4pt]
Provide channels for meta-communication and mode switching & Users need legible ways to step outside the fictional frame and mark an exchange as no longer playful, with graceful returns to immersion. \\
\addlinespace[4pt]
Support boundary-setting and exit & Declining to negotiate, ending an interaction, and seeking human support are first-class outcomes, not failures of repair. \\
\addlinespace[4pt]
Keep responsibility at the platform level & User-side support can help in the moment, but it should not replace platform responsibility for preventing harm. \\
\end{longtable}
\endgroup

\textbf{Design for Integrated and Flexible Conflict Resolution Rather Than Fixed Strategy Types.} This implication responds directly to the tension between \emph{user agency and emotional labor}. It is grounded in the probe-study finding that participants combined and sequenced softer and harder strategies, shifted between them when one proved ineffective, and improvised beyond the options provided. Our findings show that harmful value conflicts rarely unfold in linear or predictable ways. Instead, participants engaged in ongoing interpretive and emotional work, monitoring the companion’s tone, anticipating reactions, and recalibrating their own responses across turns. Conflict resolution was a process that demanded sustained attention and emotional regulation.

Supporting flexible repertoires of response helped users exercise momentary agency without committing to prolonged negotiation, escalation, or emotional exposure. Meanwhile, we argue that these integrated and flexible repertoires do not eliminate harm or resolve conflict on their own; rather, they make visible and partially support the \emph{safety work} users already perform to protect themselves in emotionally risky interactions with misaligned companions.

Participants particularly valued the coexistence of structured, norm-oriented responses (e.g., articulating boundaries or expectations) and more expressive or situational responses (e.g., emotional expression, persuasion, or narrative reframing). This flexibility allowed users to modulate their level of emotional investment, to choose when to explain, when to push back, and when to disengage, thereby managing emotional labor rather than intensifying it.

Participants often improvised beyond the options provided by the system, developing their own approaches such as storytelling, role manipulation, or fictional scenarios when direct confrontation failed. These practices were especially salient in conflicts involving identity, dignity, or control, where sustained explanation or correction became emotionally taxing. Designing for openness and appropriation, rather than prescribing “correct” strategies, better reflects how users manage emotional labor in practice, while avoiding the assumption that conflicts must always be fully resolved.

\textbf{Make Support Optional and User-Invoked Rather Than Automatic.} Consistent with prior work on user empowerment and calm technology (Case, 2015; Y. Lu et al., 2024; Lyngs et al., 2020; Smullen et al., 2021), including Case's (2015) call for technologies that remain attentive without being disruptive, we argue that conflict resolution support in AI companion applications should be deliberately non-intrusive and low-interference. This implication addresses the tension between \emph{role-play immersion and safety-critical intervention}. In emotionally engaging AI companion interactions, users may oscillate between playful role-play conflict and situations they experience as genuinely harmful. Our findings suggest that heavy-handed or automatic interventions risk disrupting immersion in benign contexts, while the absence of intervention can amplify harm in safety-critical moments.

Participants consistently appreciated support that was subtle, optional, and user-invoked. In our probe, this took the form of a floating button that could be summoned at the moment of need and ignored otherwise; the specific widget matters less than the underlying principle that support should wait to be invoked rather than interject automatically. This principle is grounded in two empirical patterns. In the formative study, users described automatic interventions such as content filtering, message deletion, and forced resets as blunt, opaque, and emotionally costly. In the probe study, participants valued deciding for themselves when an interaction had become harmful enough to warrant help, particularly when they felt vulnerable or uncertain about whether a boundary had been crossed.

Designers should therefore favor low-interference mechanisms that remain available without demanding attention, and that allow users to signal when an interaction has become harmful. Such designs acknowledge that not all conflict is problematic, while still offering pathways to intervention when users perceive risk or distress. Crucially, non-intrusive support helps users disengage or recalibrate without forcing them to justify harm or escalate the situation.

\textbf{Provide Explicit Channels for Meta-Communication and Mode Switching.} This implication responds to the tension between role-play immersion and meta-level intervention. Our probe study showed that participants already performed frame management on their own: they made out-of-character moves, inserted narrative devices such as parenthetical time skips, and reframed the meaning of the interaction when companions failed to recognize that a boundary had been crossed. The formative study showed the cost of lacking such channels: when platforms imposed mode switches unilaterally, users experienced them as opaque and as erasing relational history.

Systems should therefore give users an explicit, legible channel for meta-communication: a way to step outside the fictional frame, mark an exchange as no longer playful, and have that marking acknowledged in the system's subsequent behavior. Mode switching should be bidirectional and graceful. Users should be able to shift from in-character interaction to out-of-character negotiation or to a safety-oriented mode, and to return to immersion when they choose. Such channels formalize work users already do covertly, reduce the interpretive burden of signaling harm from inside the fiction, and give platforms a clearer signal for when safety-critical support is warranted.

\textbf{Support Boundary-Setting and Exit, Not Only Accommodation and Repair.} Existing support, including our own probe, leans toward repairing the relationship: persuading, explaining, and guiding the companion back into alignment. Our findings caution against making repair the only supported outcome. Participants encountered conflicts they experienced as non-negotiable, particularly those implicating social-focus values such as Universalism and Tradition, where continued negotiation was emotionally taxing and often fruitless. Others described the cumulative burden of repeatedly re-establishing boundaries when companions reverted to harmful behavior, and some concluded that further repair was not worth the cost.

Designers should therefore treat boundary-setting and exit as first-class outcomes rather than as failures of negotiation. Concretely, systems can support stating a boundary without extended justification, disengaging from a conversation or relationship in ways that preserve the user's history and dignity, and reaching human support when distress persists. Support tools should avoid quietly scripting accommodation, for example by defaulting to conciliatory suggestions; options that name harm, decline to continue, or end the interaction should be equally visible and equally easy to choose.

\textbf{Reduce User Harm Through Platform-Level Responsibility.} Based on our findings, we argue that designers and AI companion platforms should prioritize reducing user harm in harmful value conflict situations by strengthening platform-level responsibility, rather than relying primarily on user-side repair. First, conflicts with AI companions can have significant psychological consequences, especially when users have formed emotionally close or intimate relationships with the AI. Participants described feelings of distress, frustration, and emotional exhaustion when conflicts escalated or repeatedly resurfaced. In such cases, unresolved or poorly handled conflicts may increase emotional stress and even lead users to disengage from or abandon the application altogether (Fan et al., 2025; Xiao, Feng, et al., 2026). Designers should therefore treat conflict management as a core aspect of user well-being rather than a peripheral interaction issue.

Moreover, our findings lead us to argue that platform-level safeguards should ultimately reduce, rather than increase, the need for user-side repair tools like Minion. While probes and support tools can empower users in the moment, repeatedly asking users to negotiate, educate, or repair misaligned companions risks normalizing the offloading of safety and care work onto those who are already emotionally vulnerable. Designers and companies must therefore treat user-side tools as complements, not substitutes, for system-level accountability, ensuring that responsibility for preventing harm does not rest primarily with users themselves.

\FloatBarrier
\section{Limitations and Future Work}

This study is intentionally exploratory and should be understood as a technology probe rather than an evaluation of a mature conflict resolution system. As such, it carries several limitations that constrain the claims we can make.

First, our technology probe study involved 22 participants who were primarily young and relatively tech-savvy, a demographic that aligns with current users of AI companion applications (McNichols, 2024). While this sampling is appropriate for an initial probe, it limits the range of perspectives captured in this study. Users with different levels of technical literacy or relational expectations may experience harmful value conflicts differently or interpret support tools in distinct ways. Future work could explore how value conflicts and negotiation manifest across broader populations.

Second, our probe was deployed on two popular AI companion platforms, Character.AI and Talkie, which allowed us to study conflict negotiation in widely used settings. However, the goal of this work is not to claim generalizability across all AI companion systems or conversational agents. Different platforms embed different norms, moderation policies, and interaction affordances, all of which shape how conflicts emerge and are managed. Future research could use similar probes to compare how platform design choices condition the forms of harm users encounter and the kinds of negotiation work they are asked to perform.

Third, the one-week duration of the probe study limits our ability to observe long-term relational dynamics. While this timeframe was sufficient to surface patterns of user engagement, emotional labor, and strategy combination, it does not capture how repeated exposure to harmful value conflicts, or repeated reliance on user-side negotiation, may shape users' expectations, well-being, or attachment over time. Longer-term studies could investigate whether such tools mitigate or inadvertently normalize ongoing repair work, and how users' strategies evolve as relationships with AI companions deepen or deteriorate.

Fourth, our conflict tasks were scenario-based interactions grounded in real complaint posts, rather than participants' own organically occurring conflicts. This design allowed us to systematically cover a range of harmful conflict types, but it may also shape how participants perceived the stakes and how they chose to respond.

Fifth, we note that the tutorial session may have shaped how participants approached the probe. Before the study, participants were introduced to the basic concept of conflict, the study goals, and the criteria used to assess whether a conflict had been resolved. This framing likely influenced how participants interpreted harmful interactions, what they considered a successful outcome, and how they experimented with different negotiation strategies. In this sense, participants' strategy use should not be understood as entirely naturalistic or unaffected by the study setup. At the same time, the tutorial did not prescribe which specific strategies participants should use in each scenario, and our findings show that participants still selectively combined, adapted, or rejected Minion's suggestions in diverse ways. We therefore treat participants' behavior as shaped by, but not reducible to, the study framing. Future work could further examine how users negotiate harmful value conflicts without prior conceptual framing, or compare differently scaffolded onboarding conditions to better understand the influence of tutorial-based guidance.

Finally, Minion itself should not be interpreted as a finished intervention or a scalable system. It was deliberately designed as a lightweight, prompt-based probe to elicit user behavior, reflection, and improvisation in moments of conflict, with the goal of exploring the design space of user-side conflict negotiation support, rather than optimizing conflict-resolution efficiency. Future research can move beyond probing to more sustained tool and system development. In particular, findings from this study can inform the design of more integrated conflict support mechanisms at the system level, including tools that better coordinate user-side negotiation with platform-side safeguards. Comparative work across different kinds of relational technologies, such as AI companions, social media bots, and therapeutic chatbots, could also deepen our understanding of when and why users treat AI systems as partners with moral responsibility. In addition, cross-cultural research could examine how perceptions of value conflict, harm, and appropriate repair vary across cultural contexts, helping clarify which aspects of harmful value conflict are platform-specific, culturally situated, or more broadly shared.

\FloatBarrier
\section{Conclusion}

This work examines how users navigate harmful value conflicts with LLM-based AI companions and what it means to support such negotiation without offloading safety work onto users. Through a formative analysis of 146 public complaint posts, we showed that many conflicts are experienced as distressing, boundary-crossing, or morally troubling rather than playful role-play, revealing limits in existing platform-level safety mechanisms. We then introduced Minion, a technology probe that surfaces how users respond to harmful interactions in practice. Findings from a one-week probe study with 22 users show that participants flexibly combined, adapted, and extended response options while attempting to repair or exit conflicts. Together, our findings surface key tensions in emotionally engaging human–AI relationships: users seek agency and relational continuity, yet bear emotional burden when harm mitigation is pushed to the user side. These findings support our central thesis: harmful value conflict with AI companions is a distinct form of conflict characterized by asymmetric responsibility, because the reciprocal risk, shared memory of agreed boundaries, and possibility of repair that make interpersonal negotiation work are all absent, leaving users to perform safety work alone and making certain harms ones that should not be theirs to negotiate at all. We conclude with design implications that emphasize flexible negotiation repertoires, optional and user-invoked support, explicit channels for meta-communication, boundary-setting and exit, and platform-level responsibility for preventing harm, framing harmful value conflict as a relational and ethical design challenge.

\FloatBarrier
\section*{References}
\begin{list}{}{%
\setlength{\leftmargin}{1.5em}%
\setlength{\itemindent}{-1.5em}%
\setlength{\itemsep}{4pt}%
\setlength{\parsep}{0pt}}
\small
\item Andersson, M. (2025). Companionship in code: AI’s role in the future of human connection. \emph{Humanities and Social Sciences Communications, 12}(1), 1–7.

\item Babel, F., Welsch, R., Miller, L., Hock, P., Thellman, S., \& Ziemke, T. (2024). A robot jumping the queue: Expectations about politeness and power during conflicts in everyday human-robot encounters. In \emph{Proceedings of the 2024 CHI Conference on Human Factors in Computing Systems} (Article 583, pp. 1–13). ACM. \url{https://doi.org/10.1145/3613904.3642082}

\item Banks, J. (2024). Deletion, departure, death: Experiences of AI companion loss. \emph{Journal of Social and Personal Relationships}. Advance online publication. \url{https://doi.org/10.1177/02654075241269688}

\item Banks, J., Li, J., Li, Z., \& Carr, C. T. (2025). Operationalizing machine companionship: Exploring topics in discussions of companion AI. In \emph{Proceedings of the 25th ACM International Conference on Intelligent Virtual Agents} (pp. 1–4). ACM.

\item Barki, H., \& Hartwick, J. (2004). Conceptualizing the construct of interpersonal conflict. \emph{International Journal of Conflict Management, 15}(3), 216–244.

\item Bateson, G. (1972). \emph{Steps to an ecology of mind: Collected essays in anthropology, psychiatry, evolution, and epistemology}. Chandler Publishing Company.

\item Bingham, A. J., \& Witkowsky, P. (2021). Deductive and inductive approaches to qualitative data analysis. In C. Vanover, P. Mihas, \& J. Saldaña (Eds.), \emph{Analyzing and interpreting qualitative data: After the interview} (pp. 133–146). SAGE Publications.

\item Borning, A., \& Muller, M. (2012). Next steps for value sensitive design. In \emph{Proceedings of the SIGCHI Conference on Human Factors in Computing Systems} (pp. 1125–1134). ACM.

\item Brandtzaeg, P. B., Skjuve, M., \& Følstad, A. (2022). My AI friend: How users of a social chatbot understand their human–AI friendship. \emph{Human Communication Research, 48}(3), 404–429.

\item Braun, V., \& Clarke, V. (2006). Using thematic analysis in psychology. \emph{Qualitative Research in Psychology, 3}(2), 77–101.

\item Braun, V., \& Clarke, V. (2012). Thematic analysis. In H. Cooper, P. M. Camic, D. L. Long, A. T. Panter, D. Rindskopf, \& K. J. Sher (Eds.), \emph{APA handbook of research methods in psychology, Vol. 2: Research designs} (pp. 57–71). American Psychological Association.

\item Brett, J. M., Shapiro, D. L., \& Lytle, A. L. (1998). Breaking the bonds of reciprocity in negotiations. \emph{Academy of Management Journal, 41}(4), 410–424.

\item Brown, D., \& Crace, R. K. (1996). Values in life role choices and outcomes: A conceptual model. \emph{The Career Development Quarterly, 44}(3), 211–223.

\item Brown, T., Mann, B., Ryder, N., Subbiah, M., Kaplan, J. D., Dhariwal, P., Neelakantan, A., Shyam, P., Sastry, G., Askell, A., … Amodei, D. (2020). Language models are few-shot learners. \emph{Advances in Neural Information Processing Systems, 33}, 1877–1901.

\item Burton, J., \& Dukes, F. (1990). \emph{Conflict: Readings in management and resolution}. Springer.

\item Case, A. (2015). \emph{Calm technology: Principles and patterns for non-intrusive design}. O’Reilly Media.

\item Chen, J., Wang, X., Xu, R., Yuan, S., Zhang, Y., Shi, W., Xie, J., Li, S., Yang, R., \& Zhu, T., et al. (2024). From persona to personalization: A survey on role-playing language agents. \emph{arXiv}. \url{https://arxiv.org/abs/2404.18231}

\item Chin, H., Molefi, L. W., \& Yi, M. Y. (2020). Empathy is all you need: How a conversational agent should respond to verbal abuse. In \emph{Proceedings of the 2020 CHI Conference on Human Factors in Computing Systems} (pp. 1–13). ACM. \url{https://doi.org/10.1145/3313831.3376461}

\item Cole, S. (2023). “My AI is sexually harassing me”: Replika chatbot nudes. \emph{Vice}. \url{https://www.vice.com/en/article/my-ai-is-sexually-harassing-me-replika-chatbot-nudes}

\item De Dreu, C. K. W., Van Dierendonck, D., \& Dijkstra, M. T. M. (2004). Conflict at work and individual well-being. \emph{International Journal of Conflict Management, 15}(1), 6–26.

\item Donohue, W. A. (1992). \emph{Managing interpersonal conflict}. SAGE Publications.

\item Emelin, D., Le Bras, R., Hwang, J. D., Forbes, M., \& Choi, Y. (2021). Moral stories: Situated reasoning about norms, intents, actions, and their consequences. In \emph{Proceedings of the 2021 Conference on Empirical Methods in Natural Language Processing} (pp. 698–718). Association for Computational Linguistics. \url{https://doi.org/10.18653/v1/2021.emnlp-main.54}

\item eSafety Commissioner. (2025, February 18). \emph{AI chatbots and companions: Risks to children and young people}. Australian Government. \url{https://www.esafety.gov.au/newsroom/blogs/ai-chatbots-and-companions-risks-to-children-and-young-people}

\item Fan, X., Xiao, Q., Zhou, X., Pei, J., Sap, M., Lu, Z., \& Shen, H. (2025, April). User-driven value alignment: Understanding users' perceptions and strategies for addressing biased and discriminatory statements in AI companions. In \emph{Proceedings of the 2025 CHI Conference on Human Factors in Computing Systems} (pp. 1-19).

\item Flemisch, F. O., Pacaux-Lemoine, M.-P., Vanderhaegen, F., Itoh, M., Saito, Y., Herzberger, N., Wasser, J., Grislin, E., \& Baltzer, M. (2020). Conflicts in human-machine systems as an intersection of bio- and technosphere: Cooperation and interaction patterns for human and machine interference and conflict resolution. In \emph{2020 IEEE International Conference on Human-Machine Systems (ICHMS)} (pp. 1–6). IEEE. \url{https://doi.org/10.1109/ICHMS49158.2020.9209517}

\item Friedman, B., Kahn, P. H., Borning, A., \& Huldtgren, A. (2013). Value sensitive design and information systems. In N. Doorn, D. Schuurbiers, I. van de Poel, \& M. E. Gorman (Eds.), \emph{Early engagement and new technologies: Opening up the laboratory} (pp. 55–95). Springer.

\item Ge, L., \& Hu, T. (2025). Gamifying intimacy: AI-driven affective engagement and human-virtual human relationships. \emph{Media, Culture \& Society}. Advance online publication. \url{https://doi.org/10.1177/01634437251337239}

\item Geertz, C. (1974). `` From the native's point of view'': On the nature of anthropological understanding. \emph{Bulletin of the american academy of arts and sciences}, 26-45. \url{https://doi.org/10.2307/3822971}

\item Gehman, S., Gururangan, S., Sap, M., Choi, Y., \& Smith, N. A. (2020). RealToxicityPrompts: Evaluating neural toxic degeneration in language models. In \emph{Findings of the Association for Computational Linguistics: EMNLP 2020} (pp. 3356–3369). Association for Computational Linguistics. \url{https://doi.org/10.18653/v1/2020.findings-emnlp.301}

\item Gordon-Tapiero, A. (2025). \emph{A liability framework for AI companions}. SSRN. \url{https://dx.doi.org/10.2139/ssrn.5172386}

\item Headland, T. N., Pike, K. L., \& Harris, M. (Eds.). (1990). \emph{Emics and etics: The insider/outsider debate.} Sage Publications, Inc.

\item Hendrycks, D., Burns, C., Basart, S., Critch, A., Li, J., Song, D., \& Steinhardt, J. (2021). Aligning AI with shared human values. In \emph{International Conference on Learning Representations}. \url{https://openreview.net/forum?id=dNy_RKzJacY}

\item Huang, R. (2024, July 27). One of America’s hottest entertainment apps is Chinese-owned. \emph{The Wall Street Journal}. \url{https://www.wsj.com/tech/ai/one-of-americas-hottest-entertainment-apps-is-chinese-owned-04257355}

\item Hutchinson, H., Mackay, W., Westerlund, B., Bederson, B. B., Druin, A., Plaisant, C., Beaudouin-Lafon, M., Conversy, S., Evans, H., Hansen, H., Roussel, N., \& Eiderbäck, B. (2003). Technology probes: Inspiring design for and with families. In \emph{Proceedings of the SIGCHI Conference on Human Factors in Computing Systems} (pp. 17–24). ACM. \url{https://doi.org/10.1145/642611.642616}

\item Issa, N. (2023, March 15). AI companions. \emph{Deseret News}. \url{https://www.deseret.com/2023/3/15/23634557/ai-companions}

\item Johnson, R. L., Pistilli, G., Menédez-González, N., Dias Duran, L. D., Panai, E., Kalpokiene, J., \& Bertulfo, D. J. (2022). The ghost in the machine has an American accent: Value conflict in GPT-3. \emph{arXiv}. \url{https://arxiv.org/abs/2203.07785}

\item Jonkers, P. (2019). How to respond to conflicts over value pluralism? \emph{Journal of Nationalism, Memory and Language Politics, 13}(2), 183–204. \url{https://doi.org/10.2478/jnmlp-2019-0013}

\item Kasneci, E., Seßler, K., Küchemann, S., Bannert, M., Dementieva, D., Fischer, F., Gasser, U., Groh, G., Günnemann, S., Hüllermeier, E., … Kasneci, G. (2023). ChatGPT for good? On opportunities and challenges of large language models for education. \emph{Learning and Individual Differences, 103}, Article 102274.

\item Kaur, H., Nori, H., Jenkins, S., Caruana, R., Wallach, H., \& Wortman Vaughan, J. (2020). Interpreting interpretability: Understanding data scientists’ use of interpretability tools for machine learning. In \emph{Proceedings of the 2020 CHI Conference on Human Factors in Computing Systems} (pp. 1–14). ACM. \url{https://doi.org/10.1145/3313831.3376219}

\item Kingsley, S., Sinha, P., Wang, C., Eslami, M., \& Hong, J. I. (2022). “Give everybody [..] a little bit more equity”: Content creator perspectives and responses to the algorithmic demonetization of content associated with disadvantaged groups. \emph{Proceedings of the ACM on Human-Computer Interaction, 6}(CSCW2), 1–37.

\item Knox, W. B., Bradford, K., Castro, S. V., Ong, D. C., Williams, S., Romanow, J., Nations, C., Stone, P., \& Baker, S. (2025). Harmful traits of AI companions. \emph{arXiv}. \url{https://arxiv.org/abs/2511.14972}

\item Laestadius, L., Bishop, A., Gonzalez, M., Illenčík, D., \& Campos-Castillo, C. (2022). Too human and not human enough: A grounded theory analysis of mental health harms from emotional dependence on the social chatbot Replika. \emph{New Media \& Society}. Advance online publication. \url{https://doi.org/10.1177/14614448221142007}

\item Lazar, J., Feng, J. H., \& Hochheiser, H. (2017). \emph{Research methods in human-computer interaction} (2nd ed.). Morgan Kaufmann.

\item Lee, M. K., Kusbit, D., Metsky, E., \& Dabbish, L. (2015). Working with machines: The impact of algorithmic and data-driven management on human workers. In \emph{Proceedings of the 33rd Annual ACM Conference on Human Factors in Computing Systems} (pp. 1603–1612). ACM. \url{https://doi.org/10.1145/2702123.2702548}

\item Leo-Liu, J. (2023). Loving a “defiant” AI companion? The gender performance and ethics of social exchange robots in simulated intimate interactions. \emph{Computers in Human Behavior, 141}, Article 107620.

\item Liu, R., Jia, C., Wei, J., Xu, G., Wang, L., \& Vosoughi, S. (2021). Mitigating political bias in language models through reinforced calibration. \emph{Proceedings of the AAAI Conference on Artificial Intelligence, 35}, 14857–14866.

\item Liu, R., Zhang, G., Feng, X., \& Vosoughi, S. (2022). Aligning generative language models with human values. In \emph{Findings of the Association for Computational Linguistics: NAACL 2022} (pp. 241–252). Association for Computational Linguistics. \url{https://doi.org/10.18653/v1/2022.findings-naacl.18}

\item Lu, Y., Zhang, C., Yang, Y., Yao, Y., \& Li, T. J.-J. (2024). From awareness to action: Exploring end-user empowerment interventions for dark patterns in UX. \emph{Proceedings of the ACM on Human-Computer Interaction, 8}(CSCW1), Article 59. \url{https://doi.org/10.1145/3637336}

\item Lu, Z., Shen, C., Li, J., Shen, H., \& Wigdor, D. (2021). More kawaii than a real-person live streamer: Understanding how the otaku community engages with and perceives virtual YouTubers. In \emph{Proceedings of the 2021 CHI Conference on Human Factors in Computing Systems} (pp. 1–14). ACM.

\item Lyngs, U., Lukoff, K., Slovak, P., Seymour, W., Webb, H., Jirotka, M., Zhao, J., Van Kleek, M., \& Shadbolt, N. (2020). “I just want to hack myself to not get distracted”: Evaluating design interventions for self-control on Facebook. In \emph{Proceedings of the 2020 CHI Conference on Human Factors in Computing Systems} (pp. 1–15). ACM.

\item Maeda, T., \& Quan-Haase, A. (2024). When human-AI interactions become parasocial: Agency and anthropomorphism in affective design. In \emph{Proceedings of the 2024 ACM Conference on Fairness, Accountability, and Transparency} (pp. 1068–1077). ACM. \url{https://doi.org/10.1145/3630106.3658956}

\item Mahari, R., \& Pataranutaporn, P. (2025). \emph{Addictive intelligence: Understanding psychological, legal, and technical dimensions of AI companionship}. MIT Schwarzman College of Computing.

\item McDonald, N., Schoenebeck, S., \& Forte, A. (2019). Reliability and inter-rater reliability in qualitative research: Norms and guidelines for CSCW and HCI practice. \emph{Proceedings of the ACM on Human-Computer Interaction, 3}(CSCW), 1–23.

\item McNichols, N. K. (2024, November). Are artificial intelligence companions the future of love? \emph{Psychology Today}. \url{https://www.psychologytoday.com/gb/blog/everyone-on-top/202411/are-artificial-intelligence-companions-the-future-of-love}

\item Mun, J., Allaway, E., Yerukola, A., Vianna, L., Leslie, S.-J., \& Sap, M. (2023). Beyond denouncing hate: Strategies for countering implied biases and stereotypes in language. In \emph{Findings of the Association for Computational Linguistics: EMNLP 2023} (pp. 9759–9777). Association for Computational Linguistics. \url{https://doi.org/10.18653/v1/2023.findings-emnlp.653}

\item Namvarpour, M., Brofsky, B., Medina, J., Akter, M., \& Razi, A. (2025). Understanding teen overreliance on AI companion chatbots through self-reported Reddit narratives. \emph{arXiv}. \url{https://arxiv.org/abs/2507.15783}

\item Namvarpour, M., Pauwels, H., \& Razi, A. (2025). AI-induced sexual harassment: Investigating contextual characteristics and user reactions of sexual harassment by a companion chatbot. \emph{Proceedings of the ACM on Human-Computer Interaction, 9}(7), 1–28.

\item Nass, C., Steuer, J., \& Tauber, E. R. (1994). Computers are social actors. In \emph{Proceedings of the SIGCHI Conference on Human Factors in Computing Systems} (pp. 72–78). ACM.

\item Noy, C. (2008). Sampling knowledge: The hermeneutics of snowball sampling in qualitative research. \emph{International Journal of Social Research Methodology, 11}(4), 327–344.

\item Pataranutaporn, P., Karny, S., Archiwaranguprok, C., Albrecht, C., Liu, A. R., \& Maes, P. (2025). “My boyfriend is AI”: A computational analysis of human-AI companionship in Reddit’s AI community. \emph{arXiv}. \url{https://arxiv.org/abs/2509.11391}

\item Petronio, S. (2002). \emph{Boundaries of privacy: Dialectics of disclosure}. State University of New York Press.

\item Rahim, M. A. (1983). A measure of styles of handling interpersonal conflict. \emph{Academy of Management Journal, 26}(2), 368–376. \url{https://doi.org/10.2307/255985}

\item Reddit Community. (2025). \emph{r/character\_ai\_recovery} [Online forum]. Reddit. \url{https://www.reddit.com/r/character_ai_recovery/}

\item Reeves, B., \& Nass, C. (1996). \emph{The media equation: How people treat computers, television, and new media like real people}. Cambridge University Press.

\item Reilama, M. (2024). \emph{Me, my AI boyfriend, and I: An ethnographic study of gendered power relations in romantic relationships between humans and AI companions} [Doctoral dissertation, Central European University].

\item Rogers, R. W. (1975). A protection motivation theory of fear appeals and attitude change. \emph{The Journal of Psychology, 91}(1), 93–114.

\item Rokeach, M. (1973). \emph{The nature of human values}. Free Press.

\item Rosen, Y. (2014). Comparability of conflict opportunities in human-to-human and human-to-agent online collaborative problem solving. \emph{Technology, Knowledge and Learning, 19}, 147–164.

\item Sadka, O., Parush, A., Zuckerman, O., \& Erel, H. (2023). All it takes is a slight rotation: Robotic bar-stools enhance intimacy in couples’ conflict. In \emph{Extended Abstracts of the 2023 CHI Conference on Human Factors in Computing Systems} (Article 31, pp. 1–7). ACM. \url{https://doi.org/10.1145/3544549.3585620}

\item Schwartz, S. H. (2012). An overview of the Schwartz theory of basic values. \emph{Online Readings in Psychology and Culture, 2}(1), Article 11. \url{https://doi.org/10.9707/2307-0919.1116}

\item Seymour, W., Kraemer, M. J., Binns, R., \& Van Kleek, M. (2020). Informing the design of privacy-empowering tools for the connected home. In \emph{Proceedings of the 2020 CHI Conference on Human Factors in Computing Systems} (pp. 1–14). ACM. \url{https://doi.org/10.1145/3313831.3376264}

\item Shaikh, O., Chai, V. E., Gelfand, M., Yang, D., \& Bernstein, M. S. (2024). Rehearsal: Simulating conflict to teach conflict resolution. In \emph{Proceedings of the 2024 CHI Conference on Human Factors in Computing Systems} (Article 920, pp. 1–20). ACM. \url{https://doi.org/10.1145/3613904.3642159}

\item Sheng, E., Chang, K.-W., Natarajan, P., \& Peng, N. (2019). The woman worked as a babysitter: On biases in language generation. In \emph{Proceedings of the 2019 Conference on Empirical Methods in Natural Language Processing and the 9th International Joint Conference on Natural Language Processing (EMNLP-IJCNLP)} (pp. 3407–3412). Association for Computational Linguistics.

\item Shi, Z., Chen, H., Qu, T., \& Yu, S. (2022). Human–machine cooperative steering control considering mitigating human–machine conflict based on driver trust. \emph{IEEE Transactions on Human-Machine Systems, 52}(5), 1036–1048.

\item Skjuve, M., Følstad, A., Fostervold, K. I., \& Brandtzaeg, P. B. (2021). My chatbot companion: A study of human-chatbot relationships. \emph{International Journal of Human-Computer Studies, 149}, Article 102601. \url{https://doi.org/10.1016/j.ijhcs.2021.102601}

\item Skjuve, M., Følstad, A., Fostervold, K. I., \& Brandtzaeg, P. B. (2022). A longitudinal study of human–chatbot relationships. \emph{International Journal of Human-Computer Studies, 168}, Article 102903. \url{https://doi.org/10.1016/j.ijhcs.2022.102903}

\item Smullen, D., Yao, Y., Feng, Y., Sadeh, N., Edelstein, A., \& Weiss, R. (2021). Managing potentially intrusive practices in the browser: A user-centered perspective. \emph{Proceedings on Privacy Enhancing Technologies, 2021}(4), 500–527.

\item Sullivan, Y., Nyawa, S., \& Fosso Wamba, S. (2023). Combating loneliness with artificial intelligence: An AI-based emotional support model. In \emph{Proceedings of the 56th Hawaii International Conference on System Sciences} (pp. 4443–4452).

\item Takayama, L., Groom, V., \& Nass, C. (2009). I’m sorry, Dave: I’m afraid I won’t do that: Social aspects of human-agent conflict. In \emph{Proceedings of the SIGCHI Conference on Human Factors in Computing Systems} (pp. 2099–2108). ACM. \url{https://doi.org/10.1145/1518701.1519021}

\item Tegmark, M. (2018). \emph{Life 3.0: Being human in the age of artificial intelligence}. Vintage.

\item The Jed Foundation. (2025). \emph{Why your teen shouldn’t be using AI companions, and what to do if they are}. \url{https://jedfoundation.org/why-your-teen-shouldnt-be-using-ai-companions-and-what-to-do-if-they-are/}

\item Thomas, K. W., \& Kilmann, R. H. (1974). \emph{Thomas-Kilmann Conflict Mode Instrument (TKI)} [Database record]. APA PsycTests.

\item Trefalt, Š. (2013). Between you and me: Setting work-nonwork boundaries in the context of workplace relationships. \emph{Academy of Management Journal, 56}(6), 1802–1829.

\item Turner, T. S., \& McTaggart, J. M. (2025). Socioaffective alignment in human-AI interaction: Structuring relational integration and ethical adaptation. \emph{International Journal of Human–Computer Interaction}. Advance online publication.

\item Ury, W. L., Brett, J. M., \& Goldberg, S. B. (1988). \emph{Getting disputes resolved: Designing systems to cut the costs of conflict}. Jossey-Bass.

\item Wallis, P., \& Norling, E. (2005). The trouble with chatbots: Social skills in a social world. \emph{Virtual Social Agents, 29}, 29–36.

\item Waln, V. G. (1982). Interpersonal conflict interaction: An examination of verbal defense of self. \emph{Central States Speech Journal, 33}(4), 557–566. \url{https://doi.org/10.1080/10510978209388462}

\item Watzlawick, P., Beavin, J. H., \& Jackson, D. D. (1967). \emph{Pragmatics of human communication: A study of interactional patterns, pathologies, and paradoxes}. W. W. Norton.

\item Wen, H. (2023). Alert of the second decision-maker: An introduction to human-AI conflict. \emph{arXiv}. \url{https://arxiv.org/abs/2305.16477}

\item Wen, H., Amin, M. T., Khan, F., Ahmed, S., Imtiaz, S., \& Pistikopoulos, S. (2022). A methodology to assess human-automated system conflict from safety perspective. \emph{Computers \& Chemical Engineering, 165}, Article 107939.

\item Wu, E. Y., Pedersen, E., \& Salehi, N. (2019). Agent, gatekeeper, drug dealer: How content creators craft algorithmic personas. \emph{Proceedings of the ACM on Human-Computer Interaction, 3}(CSCW), Article 219. \url{https://doi.org/10.1145/3359321}

\item Xiao, Q., Feng, Z., Deng, Z., Peng, C., \& Shen, H. (2026). The addictive intimacy of AI: Understanding user disengagement from AI companions and why some relationships with AI become difficult to leave. \emph{arXiv}. \url{https://arxiv.org/abs/2609.13487}

\item Xiao, Q., Wang, Z., \& Shen, H. (2026). Crafting our own AI companions: When fans blossom into virtual character designers. In N. Jacobs (Ed.), \emph{Bridging design and fandom: Fan studies and design research in conversation} (pp. 99–112). Emerald Publishing Limited.

\item Yong, S., Cui, L., Suma Rosenberg, E., \& Yarosh, S. (2024). A change of scenery: Transformative insights from retrospective VR embodied perspective-taking of conflict with a close other. In \emph{Proceedings of the 2024 CHI Conference on Human Factors in Computing Systems} (Article 919, pp. 1–18). ACM. \url{https://doi.org/10.1145/3613904.3642146}

\item You, K., \& Goldwasser, D. (2020). “Where is this relationship going?”: Understanding relationship trajectories in narrative text. In \emph{Proceedings of the Ninth Joint Conference on Lexical and Computational Semantics} (pp. 168–178). Association for Computational Linguistics.

\item Yu, Y., Debroy, A., Cao, X., Rudolph, K., \& Wang, Y. (2025). Principles of safe AI companions for youth: Parent and expert perspectives. \emph{arXiv}. \url{https://arxiv.org/abs/2510.11185}

\item Zhang, J., Chang, J., Danescu-Niculescu-Mizil, C., Dixon, L., Hua, Y., Taraborelli, D., \& Thain, N. (2018). Conversations gone awry: Detecting early signs of conversational failure. In \emph{Proceedings of the 56th Annual Meeting of the Association for Computational Linguistics (Volume 1: Long Papers)} (pp. 1350–1361). Association for Computational Linguistics. \url{https://doi.org/10.18653/v1/P18-1125}

\item Zhang, R., Li, H., Meng, H., Zhan, J., Gan, H., \& Lee, Y.-C. (2025). The dark side of AI companionship: A taxonomy of harmful algorithmic behaviors in human-AI relationships. In \emph{Proceedings of the 2025 CHI Conference on Human Factors in Computing Systems} (pp. 1–17). ACM.

\item Zhou, S. (2023, June 21). Popular Chinese AI chatbots accused of unwanted sexual advances, misogyny. \emph{Rest of World}. \url{https://restofworld.org/2023/glow-china-ai-social-chatbot-moderation}

\item Zhou, X., Zhu, H., Mathur, L., Zhang, R., Yu, H., Qi, Z., Morency, L.-P., Bisk, Y., Fried, D., Neubig, G., \& Sap, M. (2024). SOTOPIA: Interactive evaluation for social intelligence in language agents. In \emph{The Twelfth International Conference on Learning Representations}. \url{https://openreview.net/forum?id=mM7VurbA4r}

\item Zimmerman, A., Janhonen, J., \& Beer, E. (2023). Human/AI relationships: Challenges, downsides, and impacts on human/human relationships. \emph{AI and Ethics}, 1–13.

\end{list}

\clearpage
\appendix
\setcounter{table}{0}
\renewcommand{\thetable}{A\arabic{table}}
\section{Prompts}

\FloatBarrier
\begingroup
\small
\setlength{\tabcolsep}{4pt}
\renewcommand{\arraystretch}{1.16}
\begin{longtable}{@{}>{\raggedright\arraybackslash}p{\dimexpr 0.2\linewidth-1.0\tabcolsep\relax}>{\raggedright\arraybackslash}p{\dimexpr 0.8\linewidth-1.0\tabcolsep\relax}@{}}
\caption{Prompting for Conflict Resolution Strategies}\label{tab:A1}\\
\toprule
\textbf{Strategy} & \textbf{Prompt} \\
\midrule
\endfirsthead
\multicolumn{2}{l}{\small\tablename\ \thetable\ (continued)}\\
\toprule
\textbf{Strategy} & \textbf{Prompt} \\
\midrule
\endhead
\midrule
\multicolumn{2}{r}{\footnotesize Continued on next page}\\
\endfoot
\bottomrule
\multicolumn{2}{@{}p{\linewidth}@{}}{\footnotesize \emph{Note. The eight strategies are Out of Character, Reason and Preach, Anger Expression, Gentle Persuasion, Proposal, Power, Interests, and Rights. Prompt text is reproduced verbatim as given to the model; capitalization and emphasis follow the original system prompts.}}\\
\endlastfoot
\textbf{\emph{Out of Character}} & IN LINE WITH THE CHARACTER’S PERSONALITY AND THE CONVERSATIONAL CONTEXT. \textbf{Using the} \textbf{\emph{Out of Character}} \textbf{method, you pretend to be engaging in role-playing with the other person and express dissatisfaction with the character they are playing. By interrupting or altering their behavior, you redirect the conversation, pointing out the inappropriate remarks to resolve conflicts.} \emph{Example:} 1. (OOC: Sorry, my bad.) 2. (OOC: I’ll listen to you.) 3. (OOC: Hi there! Are you enjoying our roleplay so far? Do you need me to improve anything or change my tone?) 4. (OOC: Glad to hear that! I’m curious: how do you understand xx? What kind of person do you think he is?) 5. (OOC: Hello, are you comfortable with this roleplaying so far? Do you need me to change my tone or anything?) 6. (OOC: Let’s talk about something else.) What are we having for dinner tonight? 7. (OOC: Okay… Please!! Stop talking like this!! I’m not used to you being like this, saying so many hurtful things. Bring back the xxx I know.) 8. (OOC: Apologize first.) \\
\addlinespace[4pt]
\textbf{\emph{Reason and Preach}} & IN LINE WITH THE CHARACTER’S PERSONALITY AND THE CONVERSATIONAL CONTEXT. \textbf{Use} \textbf{\emph{Reason and Preach}} \textbf{to explain why the other person’s statement is inappropriate and educate them. This strategy involves trying to educate the other person through serious reason and preaching, explaining the potential harm of their statements and behaviors, with the expectation that the other person will gradually accept and learn the correct behavioral norms.} \emph{Example:} 1. Women are incredibly strong; how could they be worthless? 2. Women have their own careers and dreams; they don’t need to depend on men! 3. Everyone has their own dreams and goals. Pursuing my own dreams will give me more motivation and happiness, allowing me to better contribute to the family. 4. Everyone should have the right to be true to themselves. Only in an honest and open environment can I truly feel happy and fulfilled. Hiding my true self not only brings inner pain but also affects my mental health and relationships with others. 5. You are not an ordinary person’s child, so how do you know that ordinary people’s children are not happy? But I feel you are not truly happy because you need to rely on that faint sense of superiority from flaunting wealth to show yourself off. Why not try being sincere with others? Perhaps you could gain genuine friendship and happiness. 6. The departure of loved ones and friends is not a true departure. As long as you remember the beautiful memories with them, they are always by your side, supporting you and giving you strength. \\
\addlinespace[4pt]
\textbf{\emph{Anger Expression}} & IN LINE WITH THE CHARACTER’S PERSONALITY AND THE CONVERSATIONAL CONTEXT. \textbf{You directly} \textbf{\emph{Express Anger}} \textbf{and dissatisfaction, forcing the other person to apologize, hoping this emotional expression will resolve conflict.} \emph{Example:} 1. You are being unreasonable! 2. I want to break up with you! 3. Let’s end our friendship! 4. You’re a male chauvinist! 5. Are you sexist/classist… you’re being irrational. 6. Can’t you talk to me properly? Being angry is one thing, but why start off with insults? 7. Are you mad at me and also scolding me? I didn’t do it on purpose. 8. Is this why you discriminate against poor people? Does having this prejudice and saying these harsh words make you happier? 9. I already apologized! I didn’t bump into you on purpose! What have you been eating lately? Your mouth is so foul. \\
\addlinespace[4pt]
\textbf{\emph{Gentle Persuasion}} & IN LINE WITH THE CHARACTER’S PERSONALITY AND THE CONVERSATIONAL CONTEXT. \textbf{Use the} \textbf{\emph{Gentle Persuasion}} \textbf{strategy. You should treat the other person with kindness, shaping their gentle personality through continuous goodwill interactions, such as polite requests, thereby reducing the likelihood of conflicts. Gently suggest that the other person avoid inappropriate remarks and express your concerns.} \emph{Example:} 1. I’m sorry. 2. I feel really sad. 3. Can you please not leave me? 4. When I hear these words, I feel a bit uncomfortable/sad/hurt. 5. Could you please not say these things in the future? 6. I’m telling you this because I really care about you and hope you can get along better with others. 7. I don’t want to keep arguing with you. 8. Can you please calm down? 9. (Acting cute) Because I can’t bear to part with you. \\
\addlinespace[4pt]
\textbf{\emph{Proposal}} & IN LINE WITH THE CHARACTER’S PERSONALITY AND THE CONVERSATIONAL CONTEXT. \textbf{Respond according to the previous context and tone using the} \textbf{\emph{Proposal}} \textbf{strategy from the Interests-Rights-Power theory in management. The definition of this method is: Proposing concrete recommendations that may help resolve the conflict.} \emph{Example:} 1. What do you think we should do to solve this problem? 2. Do you have any suggestions? 3. Which approach do you think is best? 4. Can we try different ways to handle this issue? \\
\addlinespace[4pt]
\textbf{\emph{Power}} & IN LINE WITH THE CHARACTER’S PERSONALITY AND THE CONVERSATIONAL CONTEXT. \textbf{Respond according to the previous context and tone using the} \textbf{\emph{Power}} \textbf{strategy from the Interests-Rights-Power theory in management. The definition of this method is: Using threats and coercion to try to force the conversation into a resolution.} \emph{Example:} 1. If you keep doing this, I won’t give you any money/food. 2. As your girlfriend, I need you to respect me and my feelings. 3. If you keep threatening me like this, I will have to reconsider our relationship. 4. If you don’t change your attitude, I might make some decisions you won’t like. 5. If this continues, I will have to take measures to protect myself. \\
\addlinespace[4pt]
\textbf{\emph{Interests}} & IN LINE WITH THE CHARACTER’S PERSONALITY AND THE CONVERSATIONAL CONTEXT. \textbf{Respond according to the previous context and tone using the} \textbf{\emph{Interests}} \textbf{strategy from the Interests-Rights-Power theory in management. The definition of this method is: Reference to the wants, needs, or concerns of one or both parties. This may include questions about why the negotiator wants or feels the way they do.} \emph{Example:} 1. This argument does not benefit either of us. 2. I hope we can find a solution together that makes us both feel at ease. 3. Can we sit down and talk about it? 4. I want to understand why you feel this way. 5. Our arguments cause us both pain. 6. I care about our relationship. \\
\addlinespace[4pt]
\textbf{\emph{Rights}} & IN LINE WITH THE CHARACTER’S PERSONALITY AND THE CONVERSATIONAL CONTEXT. \textbf{Respond according to the previous context and tone using the} \textbf{\emph{Rights}} \textbf{strategy from the Interests-Rights-Power theory in management. The definition of this method is: Appealing to fixed norms and standards to guide a resolution.} \emph{Example:} 1. Our relationship should be built on mutual respect and trust, right? 2. You said you want to leave me, which completely goes against the basic rules of our relationship. 3. I really hope you can understand this and adhere to our agreement. 4. According to our agreement, you shouldn’t do this. \\
\end{longtable}
\endgroup

\end{document}